\documentclass[preprint,3p,12pt]{elsarticle}

\usepackage{tcolorbox}
\def\thedemobiblio#1{\smallskip\par
 \list{}{\labelwidth 0pt \leftmargin 1em \itemindent -1em \itemsep 1pt}
 \small \parindent 0pt
 \parskip 1.5pt plus .1pt\relax
 \def\newblock{\hskip .11em plus .33em minus .07em}
 \sloppy\clubpenalty4000\widowpenalty4000
 \sfcode`\.=1000\relax}

\newcommand{\bfg}[1]{\mbox{\boldmath $#1$}}
\newcommand{\real}{\hbox{\rm I\kern-0.2emR}}
\newcommand{\bm}[1]{\mbox{\boldmath $#1$}}

\newtheorem{theorem}{Theorem}[section]
\newenvironment{proof}[1][Proof]{\begin{trivlist}
		\item[\hskip \labelsep {\bfseries #1}]}{\end{trivlist}}

\usepackage{amssymb}
\usepackage{amsmath}
\usepackage{amsfonts}
\usepackage{amsbsy}
\usepackage{color}
\usepackage{float}
\usepackage{graphicx}
\usepackage{tcolorbox}
\usepackage{caption}
\usepackage{subcaption}
\usepackage{lineno}
\newtheorem{remark}{Remark}[section]
\newtheorem{lemma}{Lemma}[section]
\newtheorem{example}{Example}[section]

\journal{arXiv}

\begin{document}

\title{\large {\bf
Peridynamic stress is the static first Piola-Kirchhoff Virial stress
}}

\author{Jun Li ${}^{\dag \S} $}
\author{Shaofan Li ${}^{\ddag}$\footnote{Email:shaofan@berkeley.edu}}
\author{Xin Lai ${}^{\dag \S}$}
\author{Lisheng Liu ${}^{\dag \S}$}

\address{
${}^{\dag}$State Key Laboratory of Advanced Technology for Materials Synthesis and Processing,
Wuhan University of Technology, Wuhan 430070, China,\\
$^{\ddag}$ Department of Civil and Environmental Engineering,
University of California, Berkeley, \\
California, 94720, USA; \\
$^{\S}$ Hubei Key Laboratory of Theory and Application of Advanced Materials Mechanics,
Wuhan University of Technology, Wuhan 430070, China;
}

\begin{abstract}
The peridynamic stress formula proposed by Lehoucq and Silling \cite{Lehoucq2008,Silling2008}
is cumbersome to be implemented in numerical computations.
Here, we show that the peridynamic stress tensor has
the exact mathematical expression as that of the first Piola-Kirchhoff static Virial stress
originated from Irving-Kirkwood-Noll formalism \cite{Irving1950,Noll1955}
through the Hardy-Murdoch procedure \cite{Hardy1982,Murdoch1983},
which offers a simple and clear expression for numerical calculations
of peridynamic stress.

Several numerical verifications have been carried out to validate the
accuracy of proposed peridynamic stress formula
in predicting the stress states
in the vicinity of the crack tip and other sources of
stress concentration.
The peridynamic stress is evaluated within the bond-based peridynamics with prototype microelastic brittle (PMB) material model.
It is found that
the PMB material model may exhibit nonlinear constitutive behaviors at large deformations.
The stress fields calculated through the
proposed peridynamic stress formula
show good agreements with finite element analysis results,
analytical solutions, and experimental data,
demonstrating
the promising potential of derived peridynamic stress formula in
simulating the stress states
of problems with discontinuities,
especially in the bond-based peridynamics.
\end{abstract}

\maketitle

\section{Introduction}

The peridynamics is a nonlocal continuum mechanics theory,
which was introduced by Silling \cite{Silling2000} in an attempt
to handle mechanical problems with discontinuities,
such as cracks and fractures.
The peridynamic theory employs spatial-integral equations without the use of spatial derivatives \cite{Lehoucq2008,Silling2000,Silling2005,Erdogan2014}, which is in contrast
to the partial differential equations used as the governing equations
in the classical continuum mechanics,
providing a general framework for problems involving discontinuities
or singularities in the deformation.
Therefore, the peridynamic theory enables to address
the spontaneous formation, propagation, branching
and coalescing of discontinuities such as cracks,
without the need for the special techniques of fracture
mechanics \cite{Bobaru2010,Bobaru2011,Bobaru2012}.

There are two types of peridynamic models: bond-based \cite{Silling2000}
and state-based \cite{Silling2007,Silling2010}.
The bond-based formulation was first proposed by Silling,
in which the interaction between material particles in the continuum
is described by a pairwise force function.

Due to its simplicity and clear physical interpretation,
the bond-based peridynamic theory has been to study a wide range of mechanical problems,
such as the fracture process of fiber-reinforced composite materials
\cite{Zhou2017,Javili2019,Mikata2019},
and fracture mode of quasi-brittle materials \cite{Yu2020}, which may have
difficulties to do by using other numerical
methods, such as finite element methods.

Nevertheless, as a nonlocal continuum mechanics, the bond-based peridynamics has difficulties
to calculate stress. To resolve this issue, Lehoucq and Silling  \cite{Lehoucq2008,Silling2008}
defined a so-called {\it peridynamic stress}. However, this peridynamic stress
is cumbersome to use, so that we cannot really use the bond-based
peridynamics for continua with general material constitutive relations.
In order to eliminate the limitations of the bond-based peridynamics,
Silling and his co-workers \cite{Silling2007,Silling2010} then developed a
state-based peridynamic formulation by introducing the concept
of peridynamic force states which contain information about peridynamic interactions.
While on the one hand, the state-based peridynamics lost the advantages of
the bond force potential function, and its damage description becomes
ad hoc and inconsistent with its material constitutive relation.
On the other hand, the bond-based peridynamics does not have
a usable stress tensor measure,
and this becomes a practical issue for describing the constitutive behaviors
of a material in terms of a stress tensor as in continuum mechanics  \cite{Silling2007}.
Even though there have been attempts to solve this problem, e.g. \cite{Foster2020},
the problem remains as a main barrier for the bond-based peridynamics
becoming an engineering analysis and design tool.

In this paper, we reexamine the peridynamic stress proposed by,
Lehoucq and Silling \cite{Lehoucq2008,Silling2008}, and we hope to find
a useful, computable, and simple formulation for the peridynamic stress.
According to \cite{Lehoucq2008,Silling2008},
the peridynamic stress was defined as follows,
\begin{equation}
\bfg{\mathfrak{P}}_{LS} ({\bf X}) :=  {1 \over 2} \int_{\mathcal{S}^2}
\int_{0}^{\infty}
\int_{0}^{\infty} (y+z)^2
 {\bf f}({\bf X} +y {\bf M}, {\bf X} - z {\bf M}) \otimes
 {\bf M} dz d y d \Omega_{\bf M},
\label{eq:Nonlocal-BLM1}
\end{equation}
where $\bfg{\mathfrak{P}}_{LS}({\bf X})$ denotes the nonlocal peridynamic stress
tensor at the material particle $\bf X$, and the subscript means that this
is the Lehoucq-Silling definition.
In the expression of Eq. (\ref{eq:Nonlocal-BLM1}),
$\mathcal{S}^2$ is the unit sphere,
$d \Omega_{\bf M}$ denotes a differential solid angle on $\mathcal{S}^2$
in the direction of any unit vector $\bf M$,
and $\bf f$ represents corresponding pairwise force density.

The derived peridynamic stress tensor is obtained from the peridynamic
bond forces that geometrically pass through the material point \cite{Lehoucq2008,Silling2008}.
Lehoucq and Silling \cite{Lehoucq2008,Silling2008,Erdogan2014} also suggested that
if the motion, constitutive model, and any non-homogeneities are sufficiently smooth,
the peridynamic stress tensor converges to a Piola-Kirchhoff stress tensor
when the horizon size converges to zero.
However, although the peridynamic stress tensor defined by
Lehoucq and Silling \cite{Lehoucq2008,Silling2008}
enables to establish a closer connection between peridynamics
and the classical view of continuum mechanics,
it is too complicated and cumbersome to be evaluated,
especially for the bond-based peridynamic model in which the pairwise
force density contains all constitutive information about materials.

To address the issue,
in this paper,
we show that in peridynamic particle formulation, which is
a special case of the nonlocal continuum,
the peridynamic stress tensor has a mathematical
expression of a weighted static Virial stress
developed by Irving and Kirkwood \cite{Irving1950} and Hardy \cite{Hardy1982}.
The outline of this paper is as follows.
The expression of peridynamic stress tensor and its derivation process
are described in Section \ref{sec:PD-stress}.
Several numerical examples are presented in Section \ref{sec:Examples}
to demonstrate the accuracy and effectiveness of the proposed peridynamic stress formulation.
Final remarks are drawn in Section \ref{sec:Conclusion}.

\section{Peridynamic stress tensor}
\label{sec:PD-stress}

In classical continuum mechanics,
the equation of motion can be expressed in a differential form with
respect to the referential configuration locally at a material particle $\bf X$ and time $t$,
\begin{equation}
\rho \ddot{\bf u} ({\bf X}, t) =  \nabla \cdot {\bf P}({\bf X},t) + {\bf b}({\bf X},t),
\label{eq:Local}
\end{equation}
where $\rho$ is the material density,
$\ddot{\bf u}$ is the acceleration vector field,
${\bf P}({\bf X},t)$ is the local first Piola-Kirchhoff stress,
the symbol $\nabla$ is the divergence operator,
and ${\bf b} ({\bf X}, t)$ is a prescribed body force density field.
As a nonlocal continuum model,
the expression of peridynamics is written in an integro-differential form
without using spatial derivatives as the stress divergence
is replaced with an integer over the peridynamic bond interactions within the family $\mathcal{H}_{X}$ \cite{Silling2000,Silling2007,Silling2010}.
Therefore, the peridynamic equation of motion is \cite{Lehoucq2008,Silling2000}:
\begin{equation}
\rho \ddot{\bf u} ({\bf X}, t) = \int_{\mathcal{B}}
 {\bf f}({\bf X}^{\prime}, {\bf X}, t)  d V_{{ X}^{\prime}} + {\bf b}({\bf X},t),~~\forall {\bf X}
 \in \mathcal{B},
\label{eq:Nonlocal-BLM2}
\end{equation}
where $\mathcal{B} \in \real^3$,
${\bf X}$  is the material particle in the referential configuration of a region $\mathcal{B}$,
$d V_{{X}^{\prime}}$ is the volume associated with material particle
${\bf X}^{\prime}$ in the referential configuration,
and $t \geq 0$ is the time. Note that in peridynamics
the nonlocal effect is characterized by a neighborhood support
$\mathcal{H}_X=\{{\bf X}^{\prime} \in \mathcal{B}:|{\bf X} - {\bf X}^{\prime}| < \delta\}$
for the material
point ${\bf X}$, which is also called as the horizon. $\delta$ is defined as the radius of
the horizon, which may be understood as the physical length
scale of the nonlocal interaction.
In the integral in the above equation,
$\bf f$ denotes the pairwise force density per unit volume square ($\rm {force/volume^2}$),
that ${\bf X}^{\prime}$ exerts on ${\bf X}$.
Conservation of linear and angular momenta requires that the force density  $\bf f$ is antisymmetric
with respect to the positions of ${\bf X}$ and ${\bf X}^{\prime}$
(see \cite{Lehoucq2008,Silling2008,Silling2010}),
\begin{equation}
{\bf f} ({\bf X}^{\prime}, {\bf X}) :=
{\bf t}({\bf  X}^{\prime}, {\bf X}) - {\bf t}({\bf X}, {\bf X}^{\prime})
= - {\bf f} ({\bf X}, {\bf X}^{\prime}),
\label{eq:Fdef}
\end{equation}
where ${\bf t}({\bf  X}^{\prime}, {\bf X})$ represents
the force state vector that material particle ${\bf  X}^{\prime}$ exerts on the material particle $\bf X$.

Although Eq. (\ref{eq:Nonlocal-BLM2}) extends
the balance equation of linear momentum to nonlocal media,
it loses some valuable properties
that are associated with the local balance law such as the divergence theorem.
Noticing such inadequacy,
Lehoucq and Silling \cite{Lehoucq2008,Silling2008} proposed the following
nonlocal {\it Peridynamic Stress Tensor},
\begin{equation}
\bfg{\mathfrak{P}}_{LS}({\bf X}) :=  {1 \over 2} \int_{\mathcal{S}^2}
\int_{0}^{\infty}
\int_{0}^{\infty} (y+z)^2
 {\bf f}({\bf X} +y {\bf M}, {\bf X} - z {\bf M}) \otimes
 {\bf M} dz d y d \Omega_{\bf M},
\label{eq:Nonlocal-BLM3}
\end{equation}
Lehoucq and Silling \cite{Lehoucq2008,Silling2008}
also demonstrated a relationship between the nonlocal
peridynamic stress tensor $\bfg{\mathfrak{P}}_{LS}$ and the pairwise force density ${\bf f}$, i.e.
\begin{equation}
\nabla \cdot \bfg{\mathfrak{P}}_{LS}({\bf X})
= \int_{\mathcal{H}_{X}} {\bf f}({\bf X}^{\prime},
{\bf X}) dV_{{X}^{\prime}},
\label{eq:Relation}
\end{equation}
As a consequence,
the peridynamic equation of motion (see Eq. \ref{eq:Nonlocal-BLM2})
is equivalent to the following partial differential equation:
\begin{equation}
\rho \ddot{\bf u} ({\bf X}, t) = \nabla \cdot \bfg{\mathfrak{P}}_{LS}({\bf X},t)
+ {\bf b}({\bf X},t),
\end{equation}
which is formally identical to the equation of motion in the classical
theory (see Eq. \ref{eq:Local}).
The peridynamic stress tensor $\bfg{\mathfrak{P}}_{LS}$ is the analogue
of the first Piola-Kirchhoff stress $\bf P$.
The proof can be found in \cite{Lehoucq2008,Silling2008},
and readers may also find discussions, interpretations, and examples in
\cite{Lehoucq2008,Silling2008}.

By doing so, we are able to relate the local divergence
to the nonlocal divergence (see Eq. \ref{eq:Relation}).
Thus, roughly speaking,
there is a general equivalence between applying the local
differential operator on the nonlocal tensor and
applying the nonlocal operator to local vector flux.
An immediate benefit of Eq. (\ref{eq:Relation}) is
that we can link the divergence of the peridynamic stress with
the boundary nonlocal linear momentum flux, i.e.
\begin{equation}
\int_{\mathcal{B}} \nabla \cdot \bfg{\mathfrak{P}}_{LS} dV_{X}
= \int_{\partial \mathcal{B}}
\bfg{\mathfrak{P}}_{LS}\cdot {\bf N} d S_{X}~,
\end{equation}
which allows us to establish peridynamic-based Galerkin weak formulations conveniently,
and maybe even formulate peridynamic theories of plates and shells.

By using Noll's lemmas \cite{Noll1955},
such nonlocal integral theorems have been late extended to a more general setting
by Gunzburger and Lehoucq \cite{Gunzburger2010} and Du et. al.
\cite{Du2013}.
In the practice of modeling and simulation, however,
the peridynamic stress defined in Eq. (\ref{eq:Nonlocal-BLM3})
is too complicated and cumbersome to be evaluated.
To resolve this issue,
in the following,
we present the main results of this work
which significantly simplifies the expression
of the peridynamic stress.

\begin{lemma}[Noll(1955)]
~~{}\\
\smallskip
Let ${\bf f} ({\bf X}, {\bf X}^{\prime})$ be a vector function defined in the initial configuration
and satisfy the following condition,
\begin{equation}
 {\bf f} ({\bf X}, {\bf X}') = - {\bf f} ({\bf X}', {\bf X}),
\end{equation}
where ${\bf X}, {\bf X}^{\prime}$ are continuous variables.
If ${\bf f} ({\bf X}^{\prime}, {\bf X})$ denotes the pairwise force density acting on ${\bf X}$
from ${\bf X}^{\prime}$.
Then we can define the nonlocal first Piola-Kirchhoff stress $\bfg{\mathfrak{P}}_{Noll} ({\bf X})$ as
\begin{equation}
\bfg{\mathfrak{P}}_{Noll} ({\bf X}) :=
-{1 \over 2} \int_{\real^3} \int_0^{1}
{\bf f} ({\bf X} + \alpha {\bf R}, {\bf X} - (1-\alpha) {\bf R})
\otimes {\bf R}  d\alpha d V_{{\bf R}}~,
\label{eq:Ptilde}
\end{equation}
which has the following property,
\begin{equation}
\nabla \cdot \bfg{\mathfrak{P}}_{Noll}
\Bigm|_{{\bf X}} = \int_{\mathcal{H}_{\bf X}}
{\bf f} ({\bf X}^{\prime}, {\bf X}) d V_{{\bf X}'}~.
\label{eq:LS-theorem}
\end{equation}
where ${\bf X} \in \Omega_0$.
\end{lemma}

This is a special case of the first Noll lemma \cite{Noll1955}.
The proof of the lemma can be found in \cite{Noll1955,Lehoucq2010}.
We think that the nonlocal first Piola-Kirchhoff stress is the same
or equivalent
as the peridynamic stress proposed by Lehoucq and Silling \cite{Lehoucq2008}
\[
\bfg{\mathfrak{P}}_{Noll} = \bfg{\mathfrak{P}}_{LS}~.
\]
In the rest of the paper, we simply denote it as $\bfg{\mathfrak{P}}$ without
distinction.
The readers may find the relevant discussions,
interpretations, and examples of the peridynamic stress in \cite{Lehoucq2008}.

\begin{figure}[H]
\centering
\includegraphics[width=4.5in]{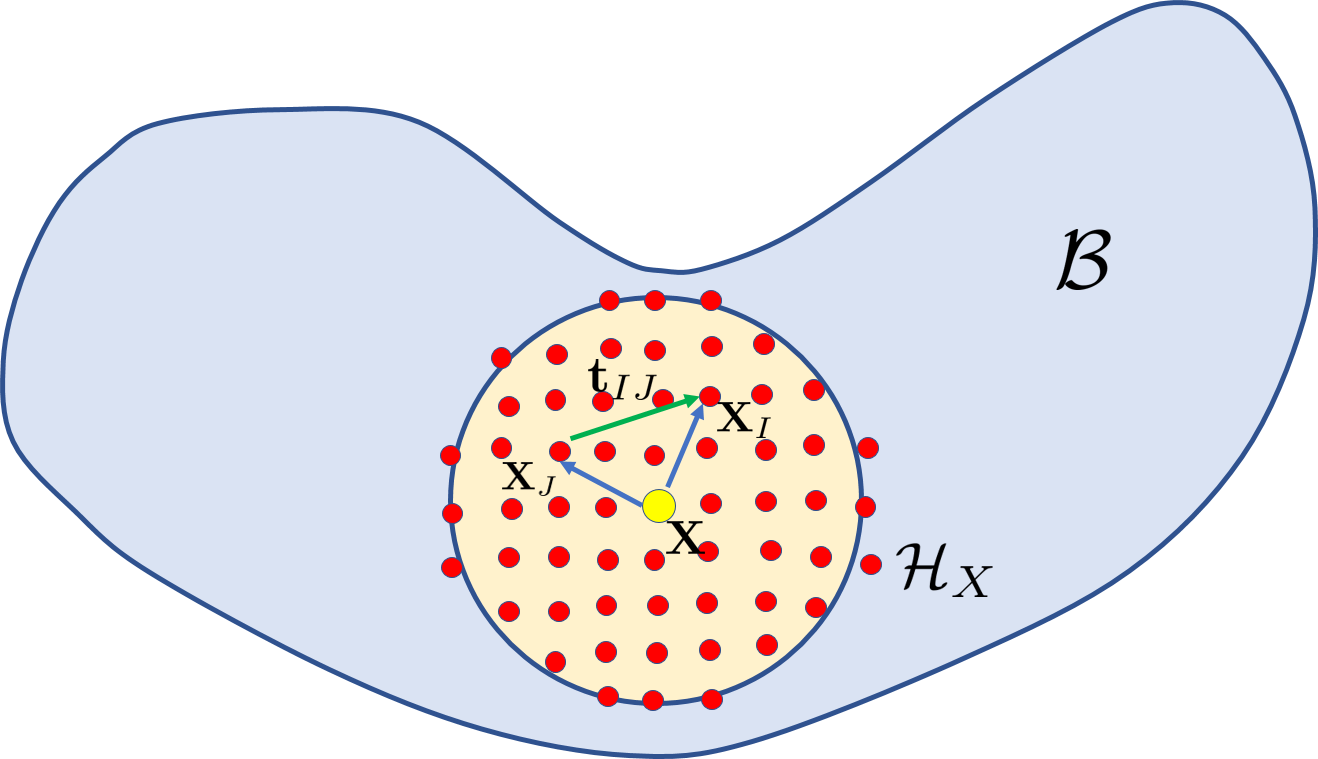}
\caption{
Illustration of peridynamic particle sampling strategy.
}
\label{fig:fig1}
\end{figure}

\begin{theorem}[Alternative form of {\it Peridynamic Stress Tensor}]
~~{}\\
\smallskip
Consider the peridynamic force density
that can be expressed as the following expression of
the Irving-Kirkwood-Hardy formulation
\cite{Silling2010,Irving1950,Hardy1982},
\begin{equation}
{\bf f}({\bf X}, {\bf X}^{\prime})
=
\sum_{I=1}^{N_X} \sum_{J=1, J\not =I}^{N_X} {\bf t}_{IJ} w ({\bf X}_I - {\bf X})
 \delta (({\bf X}_J -{\bf X}_I) - ({\bf X}^{\prime}-{\bf X})),
\end{equation}
where
${\bf X}, {\bf X}^{\prime}, {\bf X}_I$ and $ {\bf X}_J$ are material particles in the referential configuration,
${\bf t}_{IJ}$ is the force acting on the particle ${\bf X}_I$ from the particle ${\bf X}_J$ (see Fig. \ref{fig:fig1}),
$N_X$ is the total number of particles inside the horizon $\mathcal{H}_{X}$,
 $\delta ({\bf X})$
is the Dirac delta function,
and $w({\bf X}_I - {\bf X})$
is a window function or kernel function.

The nonlocal peridynamic stress defined by Noll \cite{Noll1955}
\begin{equation}
\bfg{\mathfrak{P}} ({\bf X}) :=
-{1 \over 2} \int_{\real^3} \int_0^{1}
 {\bf R} \otimes {\bf f} ({\bf X} + \alpha {\bf R}, {\bf X} - (1-\alpha) {\bf R})
 \otimes {\bf R}
 d\alpha d V_{{\bf R}}~,
\label{eq:Nonlocal-Noll}
\end{equation}
has the following exact analytical form,
\begin{equation}
\bfg{\mathfrak{P}} ({\bf X}) :=
{1 \over 2} \sum_{I=1}^{N_X} \sum_{J =1,J\not =I}^{N_X} {\bf t}_{IJ} \otimes
({\bf X}_J - {\bf X}_I) B_{IJ} ({\bf X}),~~{\bf X}_I, {\bf X}_J \in
\mathcal{H}_X,~~
\label{eq:Nonlocal-ST2}
\end{equation}
where ${\bf X}$ is the center point
of the horizon $\mathcal{H}_{X}$ and ${\bf X} \in \mathcal{B}$,
${\bf t}_{IJ} = {\bf f}({\bf X}_J, {\bf X}_I)V_I V_J$
is the force acting on the particle ${\bf X}_I$ by the particle ${\bf X}_J$,
where ${\bf X}_I, {\bf X}_J \in \mathcal{H}_{X}$, $V_I$ and $V_J$
represents the volume of material particle ${\bf X}_I$ and ${\bf X}_J$,
respectively, as shown in Fig. \ref{fig:fig1},
and
\begin{equation}
B_{IJ} ({\bf X}) = \int_0^1 w (\alpha ({\bf X}_{J} -{\bf X}_I) + {\bf X}_I - {\bf X}) d \alpha
\end{equation}
is the bond function.
\end{theorem}
\begin{proof}
$~{}$\\
Based on Noll's lemma \cite{Noll1955}, we can write the
peridynamic  first Piola-Kirchhoff stress as,
\begin{eqnarray}
\bfg{\mathfrak{P}} ({\bf X}) &=& -{1 \over 2 } \int_{\mathcal{S}^2} d \Omega_m
\int_0^{\infty} R^2 d R \int_0^1 {\bf f} ({\bf X} + \alpha R {\bf M},
{\bf X} - (1-\alpha) R {\bf M}) \otimes {\bf M} d \alpha
\nonumber
\\
&=&- {1 \over 2 } \int_{\real^3} d V_R
\int_0^1 {\bf f} ({\bf X} + \alpha  {\bf R},
{\bf X} - (1-\alpha)  {\bf R}) \otimes {\bf R} d \alpha~,
~\forall {\bf X} \in \mathcal{B}.
\label{eq:Ntilde2}
\end{eqnarray}
Considering the Hardy-Murdoch procedure
\cite{Hardy1982,Murdoch1983,Murdoch2007},
we have the following peridynamic sampling formulation (see Fig. \ref{fig:fig1})
\begin{equation}
 {\bf f} ({\bf X}^{\prime},{\bf X})
 = -\sum_{I=1}^{N_X}\sum_{J=1, J \not =I}^{N_X}{\bf t}_{IJ}
w ({\bf X}_I - {\bf X})
\delta (({\bf X}_J-{\bf X}_I) - ({\bf X}^{\prime}-{\bf X})),~~
\label{eq:Irving1a}
\end{equation}
where the window function
must satisfy the following conditions,
\begin{equation}
\int_{\mathcal{H}_{X}} w ({\bf y} - {\bf x}) d V_{ y} = 1~,
\label{eq:cond1}
\end{equation}
and
\begin{equation}
\lim_{r \to 0} w(r) ~\to~\delta (r)~.
\label{eq:cond2}
\end{equation}
The condition shown in Eq. (\ref{eq:cond1}) is the averaging requirement,
and the condition shown in Eq. (\ref{eq:cond2}) ensures
that the Dirac comb sampling can converge to
a correct continuum form of the integrand in Eq. (\ref{eq:Fdef}), i.e.
\begin{equation}
\sum_{I=1}^{N_X}\sum_{J=1, J \not =I}^{N_X}{\bf t}_{IJ}
w ({\bf X}_I - {\bf X})
\delta (({\bf X}_J-{\bf X}_I) - ({\bf X}^{\prime}-{\bf X}))
~\to~ {\bf f} ({\bf X}^{\prime}, {\bf X})= -{\bf f} ({\bf X}, {\bf X}^{\prime})~.
\end{equation}

In peridynamic model,
we often choose the following window functions:
\begin{itemize}
\item Radial step function:
\begin{equation}
w (r) = \left \{
\begin{array}{lcl}
\displaystyle {1 \over \Omega_X}, && r < \delta
\\
\\
0 ,  && {\rm otherwise}
\end{array}
\right .
\label{eq:RadialSa}
\end{equation}
where $\Omega_X = vol (\mathcal{H}_{ X}) = (4/3) \pi \delta^3$ represents the volume of horiozn $\mathcal{H}_X$, and $\delta$ is the radius of
the horizon;
\item Gaussian function;
\begin{equation}
w (r) =
\displaystyle {1 \over \delta^3 \pi^{3/2}}
\exp ( - (r/\delta)^2),
\label{eq:Gaussian}
\end{equation}
\item Cubic spline function:
\begin{equation}
w (r) = {8 \over \pi \delta^3} \left \{
\begin{array}{lcl}
\displaystyle 1 - {3 \over 2} (r/\delta)^2 + {3 \over 4} (r/\delta)^3, && r < 1/2
\\
\\
\displaystyle {1 \over 4} (2 - r/\delta)^3, &&  1/2 < r < 1
\\
\\
0 . && {\rm otherwise}
\end{array}
\right .
\end{equation}
\end{itemize}

Letting
\[
{\bf X} = {\bf X} + \alpha {\bf R},
~{\rm and}~~ {\bf X}^{\prime} = {\bf X} -(1-\alpha) {\bf R},
\]
and substituting them into Eq. (\ref{eq:Irving1a}), we then have
\begin{eqnarray}
&&
{\bf f}({\bf X},{\bf X}^{\prime})
={\bf f} ({\bf X} +\alpha {\bf R}, {\bf X} - (1-\alpha) {\bf R})
\nonumber
\\
&=& \sum_{I=1}^{N_X}\sum_{J=1, J\not =I}^{N_X} {\bf t}_{IJ}
w (({\bf X}_I - {\bf X})- \alpha {\bf R})
\delta ({\bf R} - ({\bf X}_I -{\bf X}_J)),
\label{eq:Irving2}
\end{eqnarray}
where ${\bf X}_I, {\bf X}_J \in \mathcal{H}_{X}$, and ${\bf X}_I \not = {\bf X}_J$.
Considering the following integration identities,
\begin{eqnarray}
&&\int_{-\infty}^{\infty} \delta (\xi - x) w (x-\eta) d x = w(\xi -\eta),
\end{eqnarray}
we first integrate
\begin{eqnarray}
&&\int_{\real^3}
\delta ({\bf R}- ({\bf X}_I - {\bf X}_J))
w (({\bf X}_I-{\bf X}) - \alpha {\bf R}) {\bf R} d  V_R
\nonumber
\\
&&=  ({\bf X}_I - {\bf X}_J )
w \bigl(({\bf X}_I-{\bf X}) - \alpha ({\bf X}_I - {\bf X}_J )\bigr)~.
\end{eqnarray}
Next, following Ref. \cite{Hardy1982}, we may define
the second integral as
the so-called bond function, i.e.
\begin{equation}
 B_{IJ} ({\bf X}) = \int_0^{1}
  w (\alpha({\bf X}_J-{\bf X}_I) + {\bf X}_I - {\bf X}) d \alpha.
\end{equation}
Thus, we have
\begin{eqnarray}
\bfg{\mathfrak{P}} ({\bf X}) &=& {1 \over 2 }
\bigl(
\sum_{I=1}^{N_X}\sum_{J=1, J \not = I}^{N_X} {\bf t}_{IJ}
 \otimes ({\bf X}_J - {\bf X}_I) \bigr) B_{IJ} ({\bf X})~,
\label{eq:Ntilde3}
\end{eqnarray}
which is called the Hardy stress (see \cite{Hardy1982,Zimmerman2004}).

\begin{figure}[H]
\centering
\includegraphics[width=4.0in]{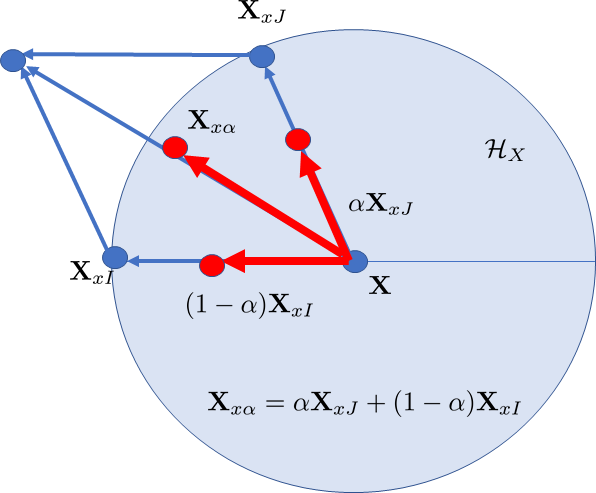}
\caption{
Graphic illustration of
the bond integration variable ${\bf X}_{x\alpha} \in \mathcal{H}_X$,
where ${\bf X}_{x\alpha} = \alpha {\bf X}_{xJ} + (1-\alpha) {\bf X}_{xI}$.
}
\label{fig:fig2}
\end{figure}

If we choose $w({\bf x})$ as the spherical radial step function (see Eq. \ref{eq:RadialS}),
one can see that
\[
w(\alpha({\bf X}_J - {\bf X}_I) + {\bf X}_I -{\bf X})
=w(\alpha({\bf X}_J-{\bf X})+ (1-\alpha) ({\bf X}_I - {\bf X}))~.
\]
If ${\bf X}_I, {\bf X}_J \in \mathcal{H}_X$, we can see that
\[
{\bf X}_{x\alpha}:= \alpha({\bf X}_J-{\bf X})+ (1-\alpha) ({\bf X}_I - {\bf X}) \in \mathcal{H}_X.
\]
This is because that
\[
|\alpha({\bf X}_J-{\bf X})+ (1-\alpha) ({\bf X}_I - {\bf X})| \leq |\alpha({\bf X}_J-{\bf X})
+ (1-\alpha) ({\bf X}_J - {\bf X})| = |{\bf X}_J - {\bf X}| \le \delta
\]
if $|{\bf X}_J - {\bf X}| \ge |{\bf X}_I - {\bf X}|$, and vice vera
\[
|\alpha({\bf X}_J-{\bf X})+ (1-\alpha) ({\bf X}_I - {\bf X})| \leq |\alpha({\bf X}_I-{\bf X})
+ (1-\alpha) ({\bf X}_I - {\bf X})| = |{\bf X}_I - {\bf X}| \le \delta
\]
if $|{\bf X}_J - {\bf X}| \le |{\bf X}_I - {\bf X}|$ as shown in Fig. \ref{fig:fig2}.
Thus, it is readily to show that
\[
B_{IJ} ({\bf X}) = {1 \over \Omega_X},~~~~{\rm if}~{\bf X}_I, {\bf X}_J \in \mathcal{H}_X
\]

For this special case, the peridynamic stress has the following expression,
\begin{tcolorbox}
\begin{eqnarray}
\bfg{\mathfrak{P}} ({\bf X}) &=& {1 \over 2 \Omega_X }
\bigl(
\sum_{I=1}^{N_X}\sum_{J=1, J \not = I}^{N_X} {\bf t}_{IJ}
 \otimes ({\bf X}_J - {\bf X}_I) \bigr) ~.
\label{eq:Ntilde4}
\end{eqnarray}
\end{tcolorbox}
The equation (\ref{eq:Ntilde4}) affirms
that the peridynamic stress is
the first Piola-Kirchhoff Virial stress.

\end{proof}

\bigskip

\begin{example}
Considering $w({\bf x})$ as the Gaussian function (see Eq. \ref{eq:Gaussian})
\[
w(r) = {1 \over \delta^3 (2\pi)^{3/2}} \exp (- {1 \over 2}(r/\delta)^2),
\]
we then have
\[
w(\alpha {\bf X}_{IJ} + {\bf X}_{I} -{\bf X} )= {1 \over \delta^3 (2\pi)^{3/2}} \exp \bigl(
- {1 \over 2 \delta^2} (\alpha^2 {X}^2_{IJ} + 2 \alpha \cos \theta X_{IJ} X_{xI} + X_{xI}^2) \bigr),
\]
where ${\bf X}_{IJ} = {\bf X}_J -{\bf X}_I$, $X_{IJ} =|{\bf X}_{IJ}|$;
${\bf X}_{xI} = {\bf X}_I - {\bf X}$ and $X_{xI}=|{\bf X}_{xI}|$,
and
\[
\cos \theta = {{{\bf X}_{IJ} \cdot {\bf X}_{xI}}
 \over {|{\bf X}_{IJ}| |{\bf X}_{xI}|}}~.
 \]
Using the formula
\[
\int {1 \over \sqrt{2 \pi}} \exp (- x^2) d x =  \Phi(x) +C,
\]
where
\[
\Phi(x)={1 \over 2}
\Bigl( 1 + {\rm erf} \bigl( {x \over \sqrt{2}} \bigr) \Bigr) ,
\]
we then have the bond function $B_{IJ} ({\bf X})$,
\begin{eqnarray}
B_{IJ} ({\bf X}) &=& \int_0^{1}
w(\alpha {\bf X}_{IJ} + {\bf X}_{I} -{\bf X} ) d \alpha
\nonumber
\\
 &=&
{1 \over \delta^3 (2\pi)^{3/2}}
\exp \Bigl(
{1 \over 2 \delta^2} \bigl( - \sin^2 \theta^2 X_{xI}^2 \bigr)
\Bigr)
{1 \over b} \Phi \bigl(
a + b \alpha \bigr) \Bigm|_0^1
\nonumber
\\
&=&
{1 \over \delta^3 (2\pi)^{3/2} b}
\exp \Bigl(
{1 \over 2 \delta^2} \bigl( - \sin^2 \theta^2 X_{xI}^2 \bigr)
\Bigr)
\Bigl( \Phi \bigl(
a + b \bigr) - \Phi (a) \Bigr),
\end{eqnarray}
where
\[
a = {\cos \theta X_{xI} \over \delta},~~{\rm and}~~b= { X_{IJ} \over \delta}~.
\]

\end{example}

\bigskip

Since all measures of stress are interrelated,
the different description of stress states
at a given material particle
within the framework of peridynamics can be found.
Define the nonlocal deformation gradient ${\bf F}$
at material point ${\bf X}$ as (see
\cite{Silling2007,Silling2010,Silling2009}),
\begin{equation}
{\bf F} ({\bf X}) =
\left \{
\int_{\mathcal{H}_{{X}}} w (|{\bf X}^{\prime} - {\bf X}|)
({\bf x}^{\prime} - {\bf x}) \otimes ({\bf X}^{\prime} - {\bf X})
d V_{{X}^{\prime}}
\right \} \cdot {\bf K}^{-1}_{X},
\label{eq:deformation-gradient}
\end{equation}
where ${\bf K}$ is the shape tensor of material particle ${\bf X}$, i.e.
\begin{equation}
{\bf K} =
\int_{\mathcal{H}_{{X}}} w (|{\bf X}^{\prime} - {\bf X}|)
({\bf X}^{\prime} - {\bf X}) \otimes ({\bf X}^{\prime} - {\bf X}) d V_{{X}^{\prime}},~
\end{equation}
and ${\bf x}= {\bf u} + {\bf X}$ is the material point
in the current configuration. Here
${\bf u}$ is the displacement field,
 ${\bf X}^{\prime} - {\bf X}$ and  ${\bf x}^{\prime} - {\bf x}$
 are the relative position
of the material points ${\bf X}^{\prime}$ and ${\bf X}$ in
the referential configuration and in the current configuration, respectively.
If we let $J = \Omega_x / \Omega_X$,
where $\Omega_x$ is the horizon volume in
the current configuration,
we can write down the peridynamic Virial stress or peridynamic
Cauchy stress as
\begin{equation}
{\bfg \varsigma} ({\bf x}) = J^{-1} \bfg{\mathfrak{P}} {\bf F}^{T} =
{1 \over 2 \Omega_x }
\bigl(
\sum_{I=1}^{N_X}\sum_{J=1, J \not = I}^{N_X} {\bf t}_{IJ}
 \otimes ({\bf x}_J - {\bf x}_I) \bigr)~,~~{\bf x}_I, {\bf x}_J \in {\mathcal{H}_x}
 \label{eq:Ntilde5}
\end{equation}

\begin{remark}
{\bf 1.}
In general, the peridynamic first Piola-Kirchhoff virial stress tensor,
\begin{eqnarray}
\bfg{\mathfrak{P}} ({\bf X}) &=&
{1 \over 2 \Omega_X }
\sum_{I=1}^{N_X}\sum_{J=1, J\not =I}^{N_X} {\bf t}_{IJ}
\otimes ({\bf X}_J - {\bf X}_I)
(B_{IJ} ({\bf X}) \Omega_X)
\label{eq:Ntilde6}
\end{eqnarray}
is a weighted static Virial stress.
\noindent
{\bf 2.} For the force density form
${\bf f}= {\bf f} ({\bf X}^{\prime}, {\bf X})$ in Eq. (\ref{eq:Nonlocal-BLM2}),
it may need a different generating function other than
that was used in the Hardy-Murdoch procedure.
For instance, one can adopt the so-called doubly-average stress
procedure \cite{Murdoch1994,Admal2010}, in which
\[
{\bf f}({\bf X}, {\bf X}^{\prime}, t) \sim
\sum_{I=1}^{N_X} \sum_{J=1. J\not =I}^{N_X} {\bf t}_{IJ}
w ({\bf X}_I - {\bf X}) w({\bf X}_J - {\bf X}^{\prime})~.
\]
However, the general conclusions of this paper will remain.
{\bf 3.}
Amazingly, the result reveals the fact that
the mesoscale peridynamic stress tensor has exactly the same expression
as that of the microscale static Virial stress,
except that it does not count for the contribution from the kinetic
energy. Moreover, the expression Eq. (\ref{eq:Ntilde6}) is so simple
that it can be readily implemented in numerical calculations without much
trouble.

\end{remark}

\section{Numerical examples}
\label{sec:Examples}
In this section, we present several numerical examples to demonstrate the accuracy and effectiveness of the proposed peridynamic stress formula
(see Eq. \ref{eq:Ntilde4})
in predicting the stress states of problems involving discontinuous deformations.
In the peridynamic stress calculations,
the force ${\bf t}_{IJ} = {\bf f}({\bf X}_J, {\bf X}_I)V_I V_J$ shown in
Eq. (\ref{eq:Ntilde4}) is computed within the framework of bond-based peridynamics, and
the radial step function is employed as the window function.

In the following, we will first give a brief review of bond-based peridynamics.
The linear microelastic is a typical example of bond-based peridynamic material models.
In this model,
each bond acts like a linear spring,
so that the pairwise force density in the bond is fully determined
by the deformation of that particular bond,
and does not depend on what happens in other bonds.
Thus,
the pairwise bond force density ${\bf f}({\bf X}^{\prime}, {\bf X})$ acting
on ${\bf X}$ from ${\bf X}^{\prime}$
can be defined as,
\begin{equation}
{\bf f}({\bf X}^{\prime}, {\bf X})
={\bf f}({\bfg \eta},{\bm \xi})
= \left \{
\begin{array}{lcl}
\displaystyle {{\bm \xi + \bm \eta} \over ||{\bm \xi +\bm \eta}||} c({\bm \xi})s, && ||{\bm \xi}|| \le \delta
\\
\\
0 , && ||{\bm \xi}|| > \delta
\end{array}
\right .
\label{eq:PD-bond-1}
\end{equation}
where ${\bm \xi} = {\bf X}^{\prime} - {\bf X}$ is the relative position,
and ${\bm \eta} = {\bf u}({\bf X}^{\prime},t) - {\bf u}({\bf X},t)$
is the relative displacement between material particles ${\bf X}^{\prime}$ and $\bf X$,
and
\begin{equation}
s = {{||\bm \eta + \bm \xi|| - ||\bm \xi||}\over ||\bm \xi||}
\end{equation}
is the bond elongation, which is therefore the change in length of a bond as it
deforms.
$c({\bm \xi})$ is called the micromodulus function
which can be evaluated by equating the energy densities of
peridynamic and classical continuum theory.
The value of $c({\bm \xi})$ is assumed as a constant at the moment, .ie.
\begin{equation}
c(\bm \xi) = c_0
= \left \{
\begin{array}{lcl}
\displaystyle {{6E} \over {\pi \delta^4(1-2\nu)}} , && {\rm 3-dimensitional~and~ plane~strain~conditions}
\\
\\
\displaystyle {{6E} \over {\pi h \delta^3(1-\nu)}} , && {\rm plane~stress~ condition}
\end{array}
\right .
\label{eq:micromodulus}
\end{equation}
where $E$ is the Young's modulus, $\nu$ is the Poisson's ratio, and $h$ is the thickness of the plane.
In the bond-based peridynamics,
the particles interact only through a pair-potential,
which leads to an effective Poisson's ratio of $1/4$ in 3-dimensional and plane strain problems and $1/3$ in plane stress problems, for an isotropic and linear microelastic material.

The simplest way to introduce failure into the linear microelastic model
in bond-based peridynamics
is by allowing bonds to break
when the corresponding stretch $s$ of the bond exceeds its critical stretch $s_0$.
Thus, to model damage,
the peridynamic force relation given in Eq. (\ref{eq:PD-bond-1})
is modified by introducing the failure parameter
$\mu (\bm \xi, t)$ to reflect bond breakage,
\begin{equation}
{\bf f}({\bf X}^{\prime}, {\bf X})
= \left \{
\begin{array}{lcl}
\displaystyle {{\bm \xi + \bm \eta} \over ||{\bm \xi +\bm \eta}||} \mu (\bm \xi, t) c({\bm \xi})s && ||{\bm \xi}|| \le \delta,
\\
\\
0  && ||{\bm \xi}|| > \delta.
\end{array}
\label{eq:PD-bond-2}
\right .
\end{equation}
The failure parameter $\mu (\bm \xi, t)$ is a history-dependent scalar-valued function that takes on values of either 1 or 0, and is expressed as \cite{Silling2005},
\begin{equation}
\mu({\bm \xi}, t)
= \left \{
\begin{array}{lcl}
\displaystyle 1, && {\rm if} ~ s({\bm \xi},t^{\prime}) < s_0 ~ {\rm for ~ all} ~ 0 < t^{\prime} < t,
\\
0 . && \rm otherwise,
\end{array}
\label{eq:failure-paramter}
\right .
\end{equation}
where $s_0$ is the critical stretch for bond failure, which is assumed as a constant at the moment.
It then leads to a notion of local damage at material particle $\bf X$,
which is defined as,
\begin{equation}
\varphi({\bf X}, t) = 1 - {\int_{\mathcal{H}_X} \mu({\bf X}^{\prime} - {\bf X}, t ) d V_{X^{\prime}} \over \int_{\mathcal{H}_X}d V_{X^{\prime}}}
\end{equation}

\begin{figure}[H]
\centering
\includegraphics[width=3.0in]{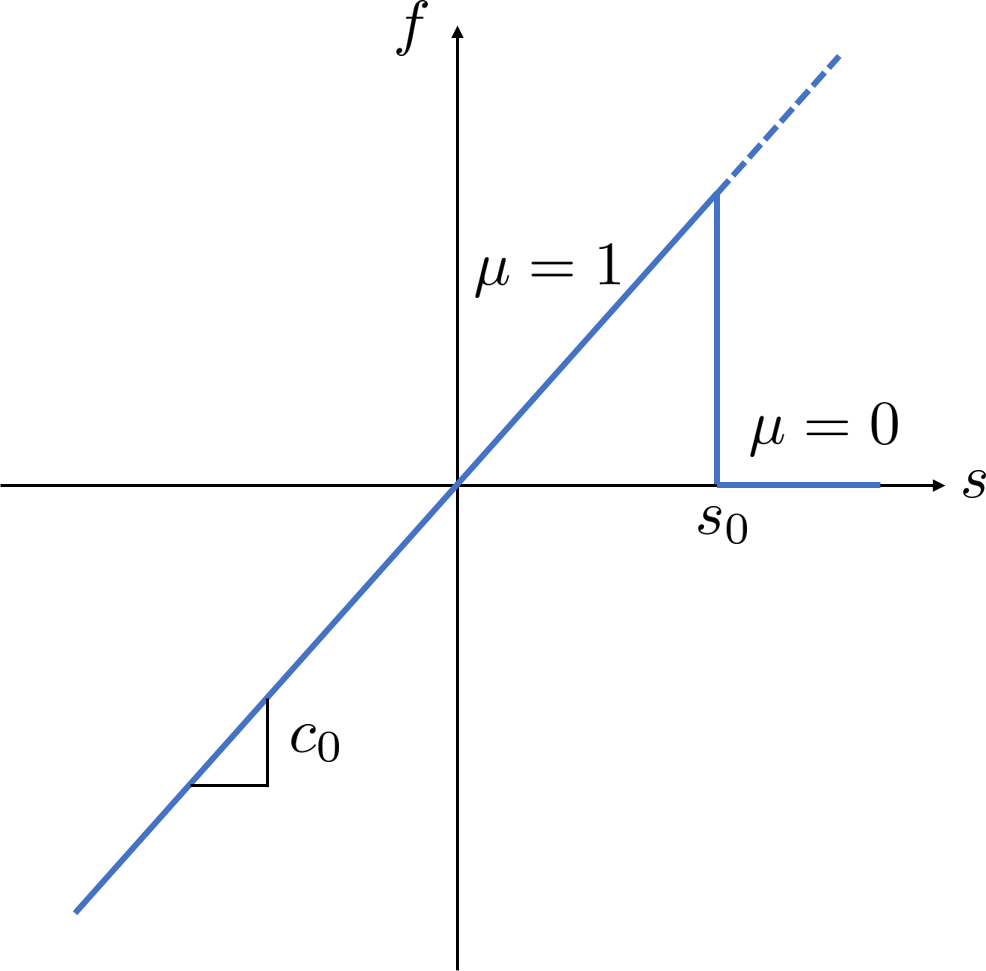}
\caption{
Constitutive relation of material particles in the PMB material model.
}
\label{fig:PMB}
\end{figure}

It is worth noting that
the linear microelastic material using Eq. (\ref{eq:micromodulus})
and combined with the bond breakage criteria given by Eqs. (\ref{eq:PD-bond-2})
and (\ref{eq:failure-paramter}) is called the {\it Prototype Micro-elastic Brittle}
(PMB) material model.
Its constitutive relation is shown in Fig. \ref{fig:PMB}.
The critical stretch $s_0$ with the PMB model can be related to the
energy release rate as derived in Silling and Askari \cite{Silling2005},
\begin{equation}
{s_0}
= \left \{
\begin{array}{lcl}
\displaystyle \sqrt{{5 G_0} \over {9 k \delta}} && {\rm three ~ dimentions},
\\
\\
\displaystyle \sqrt{{\pi G_0} \over {3 k \delta}} && \rm two ~ dimentions,
\end{array}
\label{eq:s0}
\right .
\end{equation}
where $G_0$ is the critical energy release rate of the material,
which is related to its fracture toughness, and $k$ is the bulk modulus.

In the following simulations, for verification purpose,
the calculated peridynamic stress is compared with finite element analysis results,
analytical solutions, or experimental data.
The finite element analysis is carried out using Abaqus software.
In peridynamic simulations, the adaptive dynamic relaxation (ADR) \cite{Kilic2010} has been employed for static and quasi-static problems.
In order to eliminate the surface effect,
the boundary conditions are implemented through fictitious layers
as described by Macek and Silling \cite{Macek2007},
which needs to be at least at the size of the horizon $\delta$
to ensure that the imposed boundary condition is accurately reflected in the real domain \cite{Nguyen2019}.

\subsection{Nonlinear constitutive behaviors of the PMB material model}

In general, constitutive models can provide the stress-strain relations
to describe the material responses to different loading conditions \cite{Zhang2017}.
In the bond-based peridynamics,
the PMB model is a mesoscale constitutive model,
and it assumes that the bond force and the bond stretch
are always linearly proportional before bond failure even at large deformations,
as shown in Fig. \ref{fig:PMB}.
On the other hand, Silling \cite{Silling2000} thinks that
the macroscale constitutive relation that corresponds to the PMB model
is a linear elastic model, and the correspondence between macroscale
material constants and mesoscale material constant is given in Eq. (37).

\begin{figure}[H]
\centering
\includegraphics[width=4.0in]{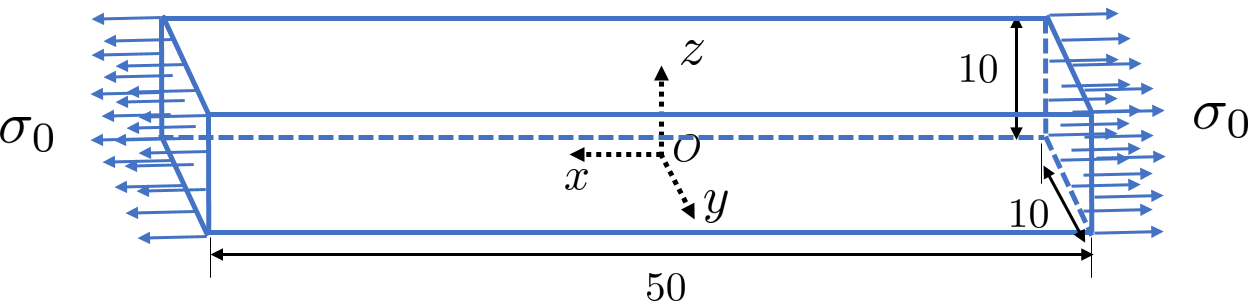}
\caption{
Graphic illustration of
a square column under uniaxial tension (unit=mm).
}
\label{fig:PMB-1}
\end{figure}

Since now we can precisely calculate the macroscale peridynamic stress
and strain, we can then find that exact macroscale stress-strain relation
that PMB model represents in continuum mechanics.
To do so, we calculated the stress-strain relation
for a square column under tensile loading
along ${\bf x}$ direction with the magnitude of
$\sigma_0=50$GPa with $c= 2.225 \times 10^{19} N/M^6$, which corresponds to
$E = 200$GPa, $\nu = 1/4$ and $\delta =1.5065 mm$ (see Fig. \ref{fig:PMB-1}).

\begin{figure}[ht]
\centering
\includegraphics[width=3.5in]{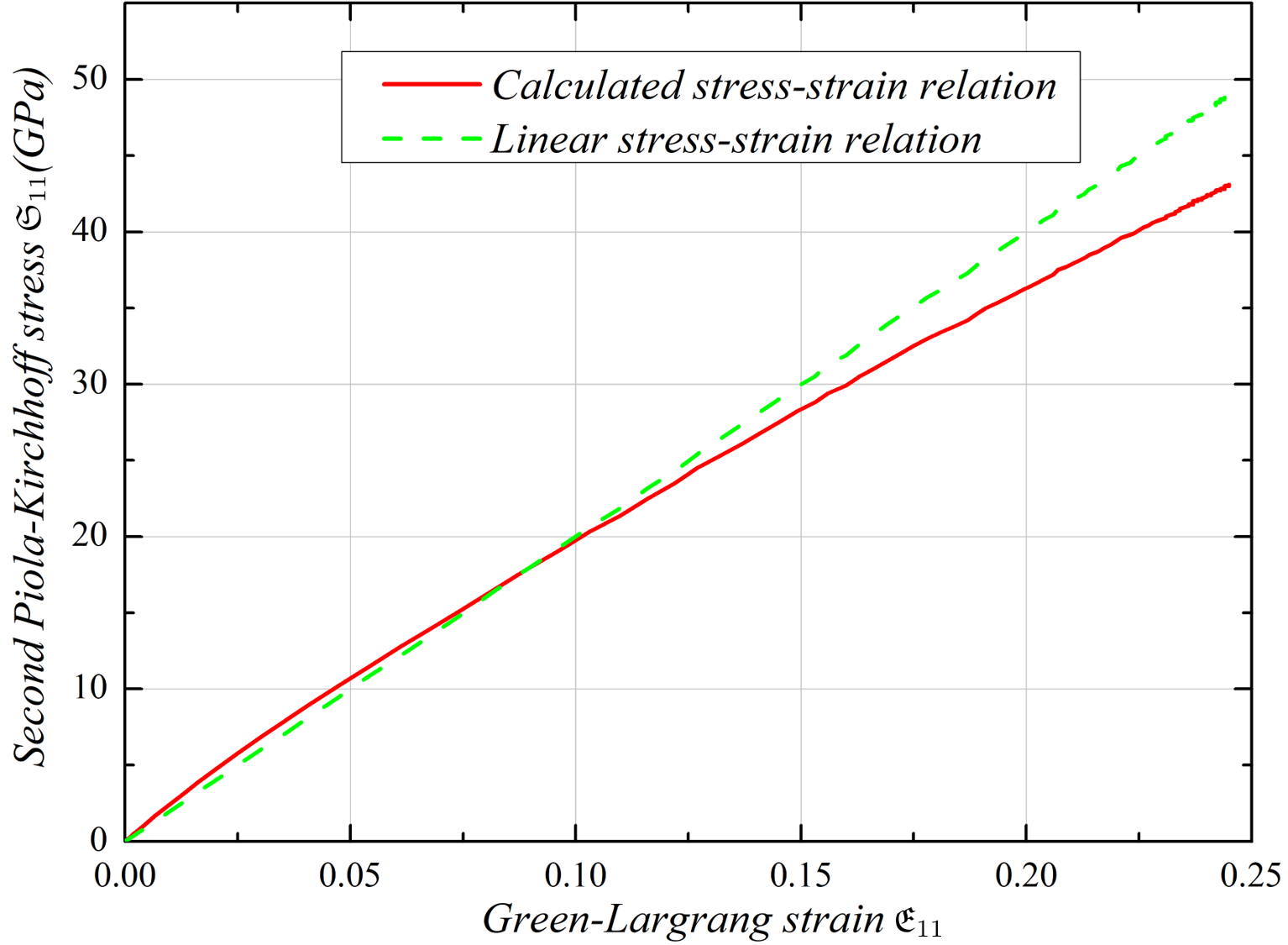}
\caption{
The calculated stress-strain relation for the square column under uniaxial tension with the PMB constitutive model.
}
\label{fig:PMB-2}
\end{figure}

Figure \ref{fig:PMB-2} shows the calculated stress-strain relation
\begin{equation}
\bfg{\mathfrak S} = \bfg{\mathfrak S} (\bfg {\mathfrak E}) ~\to~\mathfrak{S}_{11} = \mathfrak{S}_{11}( \mathfrak{E}_{11})
\label{eq:PDC}
\end{equation}
based on the PMB material constitutive model.
In Eq. (\ref{eq:PDC}),
the stress measure is the second Piola-Kirchhoff stress $\bfg{\mathfrak{S}}$
that is defined as,
\begin{equation}
\bfg{\mathfrak{S}} = {\bf F}^{-1}\bfg{\mathfrak{P}},
\end{equation}
while the strain measure is the Green-Lagrangian strain $\bfg{\mathfrak{E}}$
that is defined as,
\begin{equation}
\bfg{\mathfrak{E}} = {1 \over 2} ({\bf F}^T{\bf F} - {\bf I}^{(2)}),
\end{equation}
where ${\bf F}$ is the nonlocal deformation gradient that is defined in
Eq. (\ref{eq:deformation-gradient}).

As can be seen from Fig. \ref{fig:PMB-2},
the stress-strain relation behaves linearly at small deformations,
which is consistent with the linear elastic model
postulated by Silling \cite{Silling2000}
based on the bond-force-bond-stretch relation
of the PMB model (see Fig. \ref{fig:PMB}).
At large deformations, however,
the PMB model may exhibit nonlinear constitutive behaviors
to demonstrate material geometric nonlinearity.

To explain this, we first let
\begin{equation}
{\bfg \zeta} = {\bfg \xi} + {\bfg \eta},
\end{equation}
where ${\bm \zeta} = {\bf x}_j - {\bf x}_i$ is the relative position of material particles $j$ and $i$ in the current configuration.

Considering the Cauchy-Born rule, we assume that
in a horizon centered at $\boldsymbol{X}$, the following relation holds:
\begin{equation}
\boldsymbol{\zeta }=\boldsymbol{F}\cdot\boldsymbol{\xi}
\label{eq:general-1}
\end{equation}
where $\boldsymbol{F}$ is the deformation gradient at ${\bf X}$,
which is a constant two-point tensor
in the entire horizon.
Thus, Eq. (\ref{eq:general-1}) leads to the following equations:
\begin{eqnarray}
 \boldsymbol{F} \boldsymbol{\xi} &=&
\boldsymbol{\xi} + \boldsymbol{\eta}~~\to~~
\frac{\partial \boldsymbol{\eta}}{\partial \boldsymbol{F}}
 = \boldsymbol{I}^{(2)} \otimes \boldsymbol{\xi}
\label{eq:general-2}
\end{eqnarray}
where
\begin{equation}
\boldsymbol{I}^{(2)} \cdot\boldsymbol{\xi}=\delta_{ij}\xi_{j}\boldsymbol{E}_i
\label{eq:general-3}
\end{equation}
in which $\boldsymbol{I}^{(2)}$ is the unit second order tensor,
and $\delta_{ij}$ is the Kronecker delta.

Now we define the averaged strain energy density as follows
\begin{equation}
W(\boldsymbol{X})=\frac{1}{2\Omega_X}\int_{\mathcal{H}}\phi(\boldsymbol{\eta},\boldsymbol{\xi})
d\boldsymbol{\xi}
\label{eq:general-5}
\end{equation}
where
\begin{equation}
\phi = {1 \over 2} { c(\|{\bfg \xi} \|) s^2 \| {\bfg \xi}\| },~~{\rm where}~
s ={ \|{\bfg \eta} + {\bfg \xi} \| - \| {\bfg \xi }\| \over \| {\bfg \xi }\|}.
\end{equation}

We can then derive the first Piola-Kirchhoff stress tensor
at an arbitrary point ${\bf X}$ as
\begin{equation}
\boldsymbol{P}(\boldsymbol{X})=\frac{\partial W(\boldsymbol{X})}{\partial \boldsymbol{F}}=\frac{1}{2\Omega_X}\int_{\mathcal{H}}\frac{\partial \phi(\boldsymbol{\eta},\boldsymbol{\xi})}{\partial ||\boldsymbol{\eta}||}
\frac{\partial \boldsymbol{||\eta}||}{\partial \boldsymbol{\eta}}
\frac{\partial \boldsymbol{\eta}}{\partial \boldsymbol{F}}
d\boldsymbol{\xi}
\label{eq:general-6}
\end{equation}
where
\begin{equation}
{\partial \phi \over \partial {\bfg \eta}} = {\boldsymbol f} =c s {\bf n},~~{\rm where}~~
{\bf n} = {{\bfg \eta} + {\bfg \xi} \over \| {\bfg \eta} + {\bfg \xi}\|}
\label{eq:general-7}
\end{equation}
\begin{equation}
\frac{\partial \boldsymbol{||\eta}||}{\partial \boldsymbol{\eta}}=
\frac{\boldsymbol{\eta}}{|| \boldsymbol{\eta}||},
\label{eq:general-8}
\end{equation}
and
\begin{equation}
\frac{\partial \boldsymbol{\eta}}{\partial \boldsymbol{F}}
 = \boldsymbol{I}^{(2)} \otimes \boldsymbol{\xi}~.
\label{eq:general-9}
\end{equation}
\medskip

Substituting Eq. (\ref{eq:general-7}), (\ref{eq:general-8})
and Eq. (\ref{eq:general-9}) into Eq. (\ref{eq:general-6}),
we obtain the expression of PK-I stress as follows,
\begin{eqnarray}
\boldsymbol{P}(\boldsymbol{X})
&=&
\frac{1}{2}\int_{\mathcal{H}}
\Bigl[
\boldsymbol{f} \otimes \boldsymbol{\xi}
\Bigr]
d V_{\boldsymbol{\xi}}~
\label{eq:general-9b}
\end{eqnarray}

Consider the following peridynamic force sampling formula,
\begin{equation}
{\bf f} ({\bf X}^{\prime}, {\bf X})
= \sum_{I=1}^{N}\sum_{J=1, J \not =I}^{N}{\bf f}_{IJ}
w ({\bf X}_I - {\bf X})
\delta (({\bf X}_J-{\bf X}_I) - ({\bf X}^{\prime}-{\bf X}))~,
\label{eq:Irving1}
\end{equation}
where ${\bf f}_{IJ} = {\bf f}_J - {\bf f}_I$.

By substituting the force sampling expression in Eq. (\ref{eq:Irving1}) into
Eq. (\ref{eq:general-9}), we have
\begin{eqnarray}
{\bf P}({\bf X}) &=&
\frac{1}{2}\int_{\mathcal{H}}
\Bigl[
\boldsymbol{f} \otimes \boldsymbol{\xi}
\Bigr]
d V_{{\xi}}
\nonumber
\\
&=& \frac{1}{2} \int_{\mathcal{H}}
\sum_{I=1}^{N} \sum_{J=1, J \not =I}^{N}
w ({\bf X}_I - {\bf X})
{\bf f}_{IJ} \otimes {\bfg \xi}
\delta (({\bf X}_J-{\bf X}_I) - ({\bf X}^{\prime}-{\bf X})) d V_{{\xi}}~.
\end{eqnarray}
We choose the radial step function as the sampling function, i.e.
\begin{equation}
w (r) = \left \{
\begin{array}{lcl}
{1 \over \Omega_X}, && r < \delta
\\
\\
0 ,  && {\rm otherwise}
\end{array}
\right .
\label{eq:RadialS}
\end{equation}
where $\Omega_X = vol (\mathcal{H}_X) = (4/3) \pi \delta^3$, and $\delta$ is the radius of
the horizon.

Since ${\bf X}, {\bf X}_I \in \mathcal{H}_X$, $w({\bf X}_I - {\bf X}) =1 / \Omega_X $.
We then have the mathematical expression of the cohesive first Piola-Kirchhoff stress,
\begin{tcolorbox}
\begin{eqnarray}
{\bf P} ({\bf X}) &=& \frac{1}{2\Omega_X} \int_{\mathcal{H}}
\sum_{I=1}^{N} \sum_{J=1, J \not =I}^{N} {\bf f}_{IJ}
 \otimes {\bfg \xi}
\delta ({\bfg \xi}_{IJ} -{\bfg \xi}) d V_{{\xi}}~
\nonumber
\\
&=&  \frac{1}{2\Omega_X}
\sum_{I=1}^{N} \sum_{J=1, J \not =I}^{N} {\bf f}_{IJ}
 \otimes {\bfg \xi}_{IJ}
 \label{eq:C-stressT}
\end{eqnarray}
where ${\bfg \xi} = {\bf X}^{\prime} - {\bf X}$  and
${\bfg \xi}_{IJ} = {\bf X}_J - {\bf X}_I$
\begin{equation}
{\bf f}_{IJ} = c(||{\bfg \xi}_{IJ}||)
{||{\bfg \eta}_{IJ} +{\bfg \xi}_{IJ}|| - ||{\bfg \xi}_{IJ}|| \over
||{\bfg \xi}_{IJ}||}
{{\bfg \eta}_{IJ} + {\bfg \xi}_{IJ} \over ||{\bfg \eta}_{IJ} + {\bfg \xi}_{IJ}||}
\end{equation}
and ${\bfg \eta}_{IJ} = {\bf u}_J - {\bf u}_I$.
\end{tcolorbox}

\begin{figure}[H]
\centering
\includegraphics[width=3.5in]{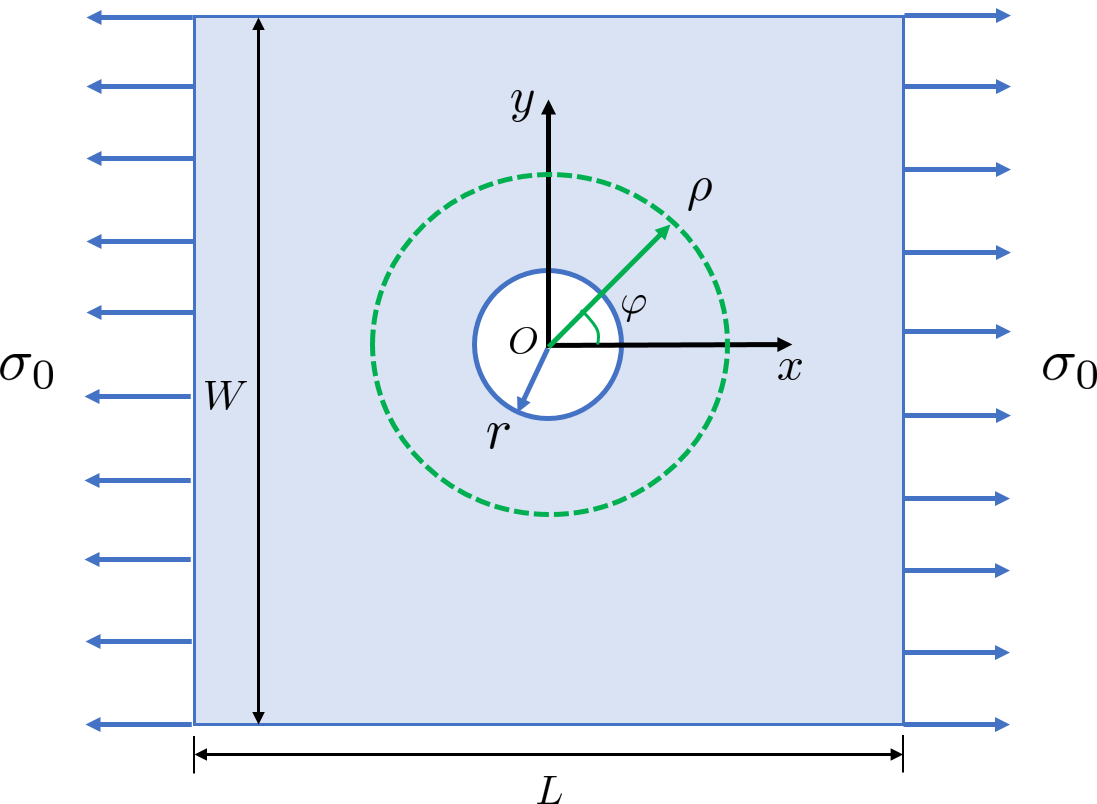}
\caption{
Graphic illustration of
the plate with a circular hole loaded under uniaxial tension.}
\label{fig:example1-1}
\end{figure}

One can see that the PK-I stress in Eq. (\ref{eq:C-stressT})
is the exact same as the nonlocal peridynamic stress in Eq. (29).

Moreover, we can then find that
\begin{equation}
{\partial^2 \phi \over \partial {\bfg \eta} \partial {\bfg \eta}}= {\partial {\bf f} \over \partial {\bfg \eta}}
=c (\|{\bfg \xi}\|) \left (
\Bigl(  { s \over \| {\bfg \eta + \bfg \xi } \|}
\Bigr) {\bf I}
+ { ({\bfg \eta} + {\bfg \xi}) \otimes ({\bfg \eta} + {\bfg \xi}) \over \| {\bfg \eta} + {\bfg \xi} \|^3 }
\right )
\label{eq:CC1}
\end{equation}
and
\begin{eqnarray}
\mathbb{C}(\boldsymbol{X})&=&\frac{\partial \boldsymbol{P}(\boldsymbol{X})}{\partial\boldsymbol{F}}
=\frac{1}{2\Omega_X}
\int_{\mathcal{H}}\frac{\partial}{\partial\boldsymbol{F}}(\frac{\partial\phi}{\partial\boldsymbol{F}})d\boldsymbol{\xi}
=\frac{1}{2\Omega_X}
\int_{\mathcal{H}}\frac{\partial}{\partial\boldsymbol{F}}(\boldsymbol{f}\otimes\boldsymbol{\xi})d\boldsymbol{\xi}
\nonumber
\\
&=&\frac{1}{2\Omega_X}\int_{\mathcal{H}}(\frac{\partial\boldsymbol{f}}{\partial\boldsymbol{F}}
\otimes\boldsymbol{\xi})d\boldsymbol{\xi}
= \frac{1}{2V}\int_{\mathcal{H}}(\frac{\partial\boldsymbol{f}}{\partial\boldsymbol{\eta}}
{\partial {\bfg \eta} \over \partial {\bf F}}
\otimes\boldsymbol{\xi})d\boldsymbol{\xi}
\nonumber
\\
&=&
\frac{1}{2\Omega_X}\int_{\mathcal{H}} \Bigl(
\frac{\partial^2 \phi}{\partial\boldsymbol{\eta} \partial {\bfg \eta}}
\otimes {\bfg \xi}
\otimes\boldsymbol{\xi} \Bigr)d\boldsymbol{\xi}
\label{eq:CC2}
\end{eqnarray}

From Eqs. (\ref{eq:CC1}) and (\ref{eq:CC2}), we can see that
the mesoscale PMB model corresponds a nonlinear macroscale
hyperelastic constitutive model.

\begin{figure}[H]
\centering
\includegraphics[width=5.0in]{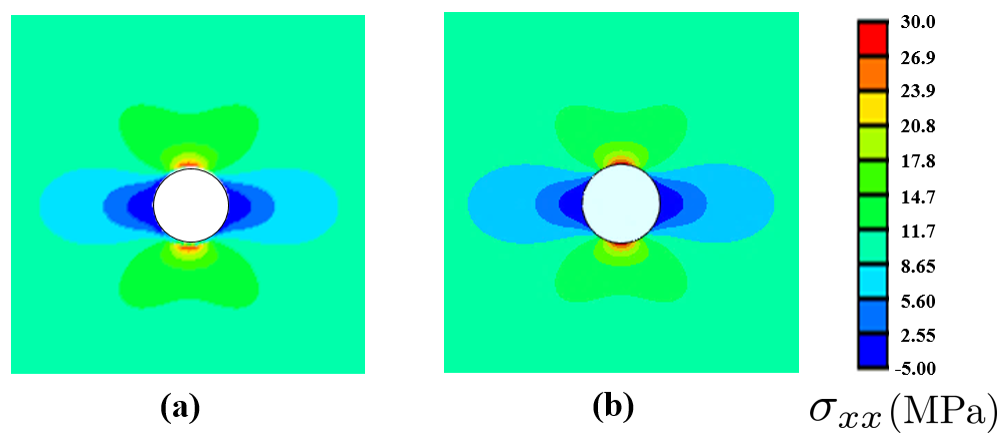}
\caption{
The normal stress $\sigma_{xx}$ for a plate with a hole under tension (a) bond-based peridynamic, (b) Abaqus results.}
\label{fig:example1-2}
\end{figure}
\subsection{A plate with a circular hole under tension}

Figure \ref{fig:example1-1} shows the square plate with dimensions
$L = W = 50$mm, and thickness $h=1$mm, where a circular hole with the diameter of $r=2.5$mm is located in the center.
The Young's modulus of the plate is $E = 192$GPa.
The plate is subjected to uniformly distributed tensile loading along $x$ direction as $\sigma_0=10$MPa.

The analytical solutions of stress components around a circular hole in an elastic infinite medium under tension are give as,
\begin{eqnarray}
\sigma_{\rho\rho} &=& {\sigma_0 \over 2}(1-{r^2 \over \rho^2})
+{\sigma_0 \over 2}(1-{r^2 \over \rho^2})(1-3{r^2 \over \rho^2})cos2\varphi
\nonumber
\\
\sigma_{\varphi\varphi} &=& {\sigma_{0} \over 2}(1+{r^2 \over \rho^2})
-{\sigma_0 \over 2}(1+3{r^4 \over \rho^4})cos2\varphi
\nonumber
\\
\sigma_{\rho\varphi} &=& \sigma_{\varphi\rho}
= -{\sigma_0 \over 2}(1-{r^2 \over \rho^2})(1+3{r^2 \over \rho^2})sin2\varphi,
\end{eqnarray}
where $\rho$ and $\varphi$ are the polar coordinates measured form the center of the circular hole (see Fig. \ref{fig:example1-1}).
The tensor measured in a polar coordinate system
can be converted into that in a Cartesian coordinate system by using tensor transformation laws that defined as,
\begin{equation}
{\bm \sigma}^{\prime} = {\bf A}{\bm \sigma} {\bf A}^{T},
\end{equation}
where $\bm \sigma^{\prime}$ and $\bm \sigma$ represents the stress tensor in
the polar coordinate system and the Cartesian system, respectively,
and $\bf A$ is the transformation matrix, which is given as ,
\begin{equation}
{\bf A} =
\begin{pmatrix}
cos\varphi & -sin\varphi \\
sin\varphi & cos\varphi
\end{pmatrix}~,
\end{equation}
for the plane stress condition.

\begin{figure}[H]
\centering
\includegraphics[width=5.0in]{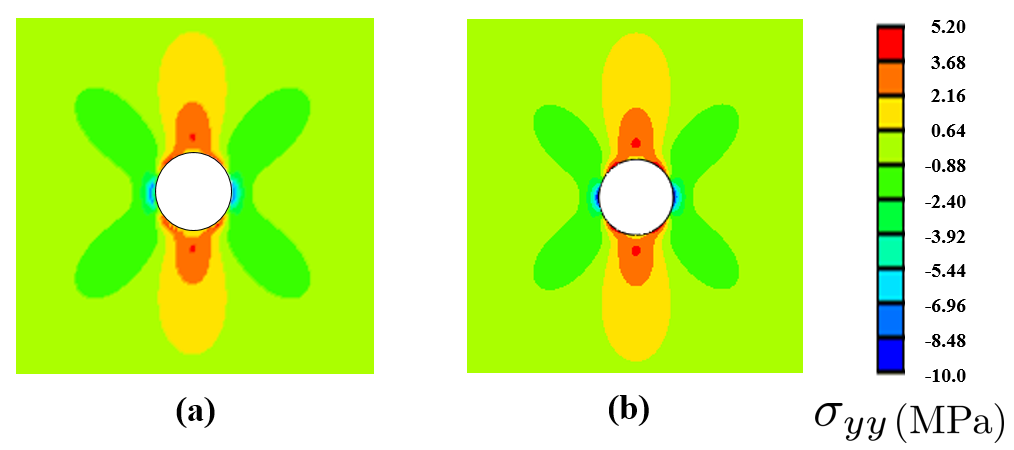}
\caption{
The normal stress $\sigma_{yy}$ for a plate with a hole under tension (a) bond-based peridynamic, (b) Abaqus results.}
\label{fig:example1-3}
\end{figure}
\begin{figure}[H]
\centering
\includegraphics[width=5.0in]{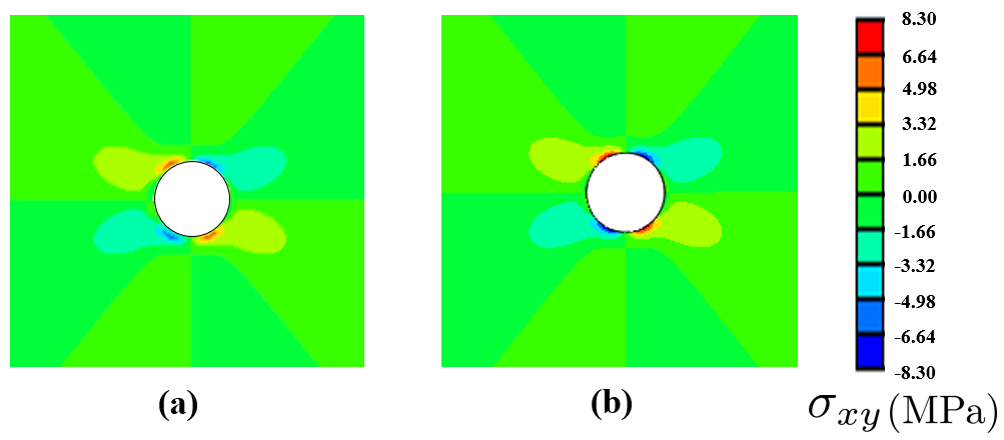}
\caption{
The shear stress $\sigma_{xy}$ for a plate with a hole under tension (a) bond-based peridynamic, (b) Abaqus results.}
\label{fig:example1-4}
\end{figure}

In the bond-based peridynamic model, the plate is uniformly discretized into a little over 2.0 million particles in a square arrangement with the grid spacing of $\Delta = 0.033$mm and the horizon radius of $\delta = 3.015\Delta$.
In this example, the accuracy of calculated peridynamic stresses is evaluated
through comparison with the finite element analysis
and analytical solutions.
In the finite element analysis, the plate with a circular hole is meshed using 248448 bilinear plane stress quadrilateral elements.

Figures \ref{fig:example1-2}-\ref{fig:example1-4} present the comparison of bond-based peridynamic and Abaqus predictions for the stress fields.
We only display the region near the central hole of the plate
for a clearer comparison.
As it can be seen from the figures,
the calculated stress distributions
using the peridynamic stress formulation (see Eq. \ref{eq:Ntilde4})
within the bond-based peridynamics
agree well with that of finite element analysis that implemented in Abaqus.

\begin{figure}[t]
\begin{minipage}{0.48\linewidth}
\begin{center}
\includegraphics[width=2.7in]{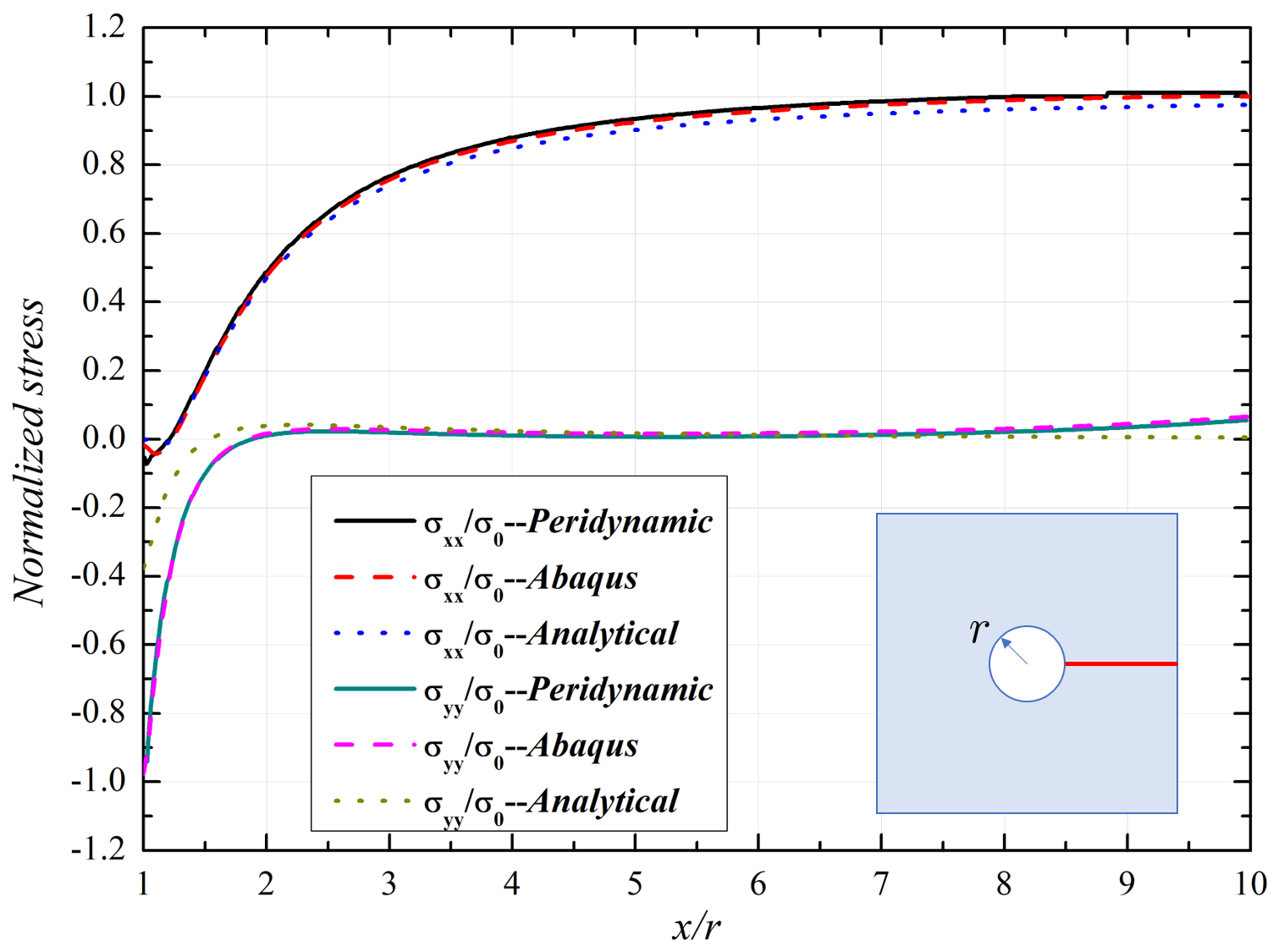}
\end{center}
\begin{center}
(a)
\end{center}
\end{minipage}
\begin{minipage}{0.48\linewidth}
\begin{center}
\includegraphics[width=2.7in]{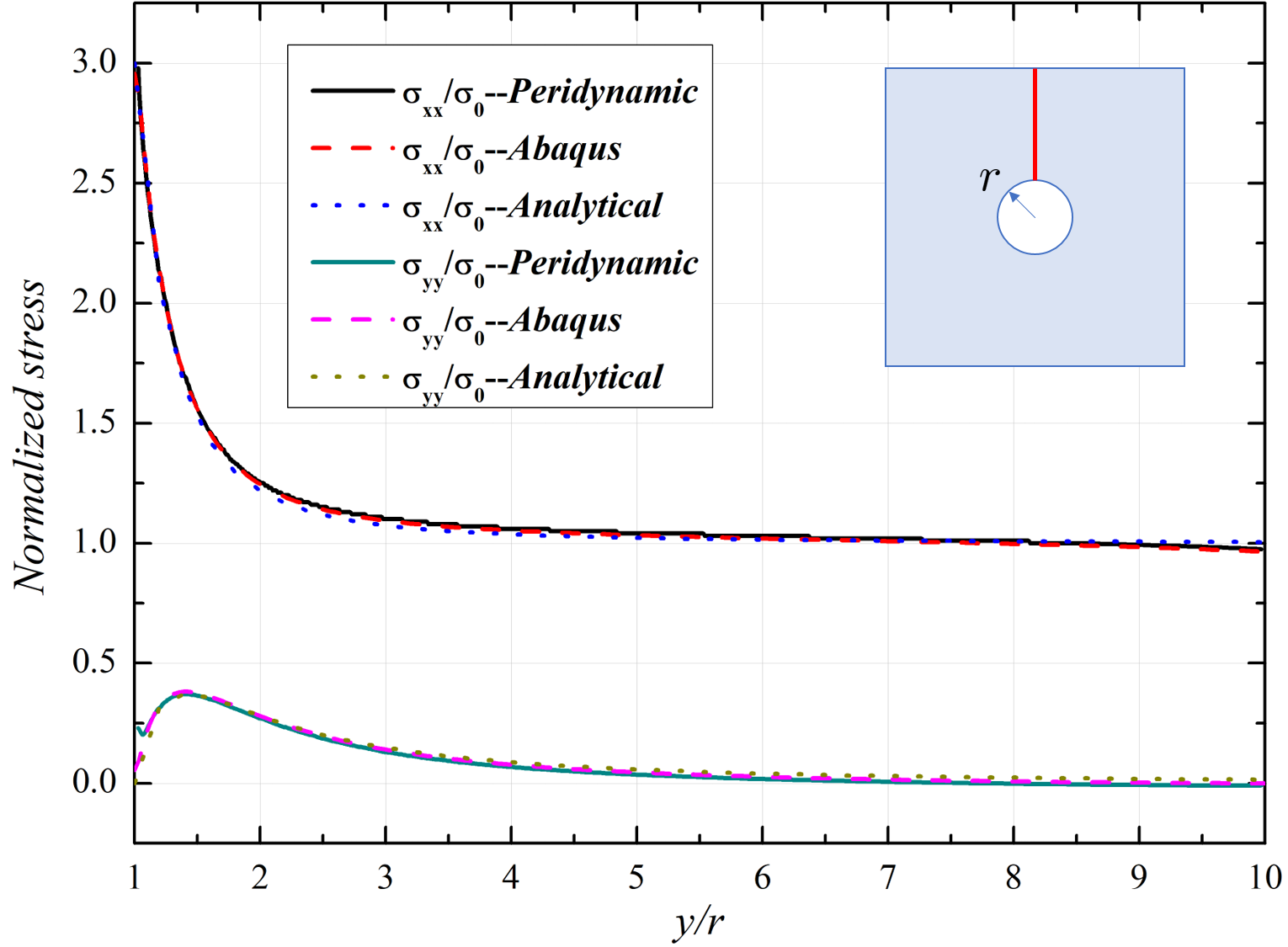}
\end{center}
\begin{center}
(b)
\end{center}
\end{minipage}
\caption{
(a)
The normalized stresses $\sigma_{xx}/\sigma_0$ and $\sigma_{yy}/\sigma_0$
along the horizontal direction shown with a red line, and
(b) The normalized stresses $\sigma_{xx}/\sigma_0$ and $\sigma_{yy}/\sigma_0$
along the vertical direction.
}
\label{fig:example1-5}
\end{figure}

\begin{figure}[H]
\centering
\includegraphics[width=4.0in]{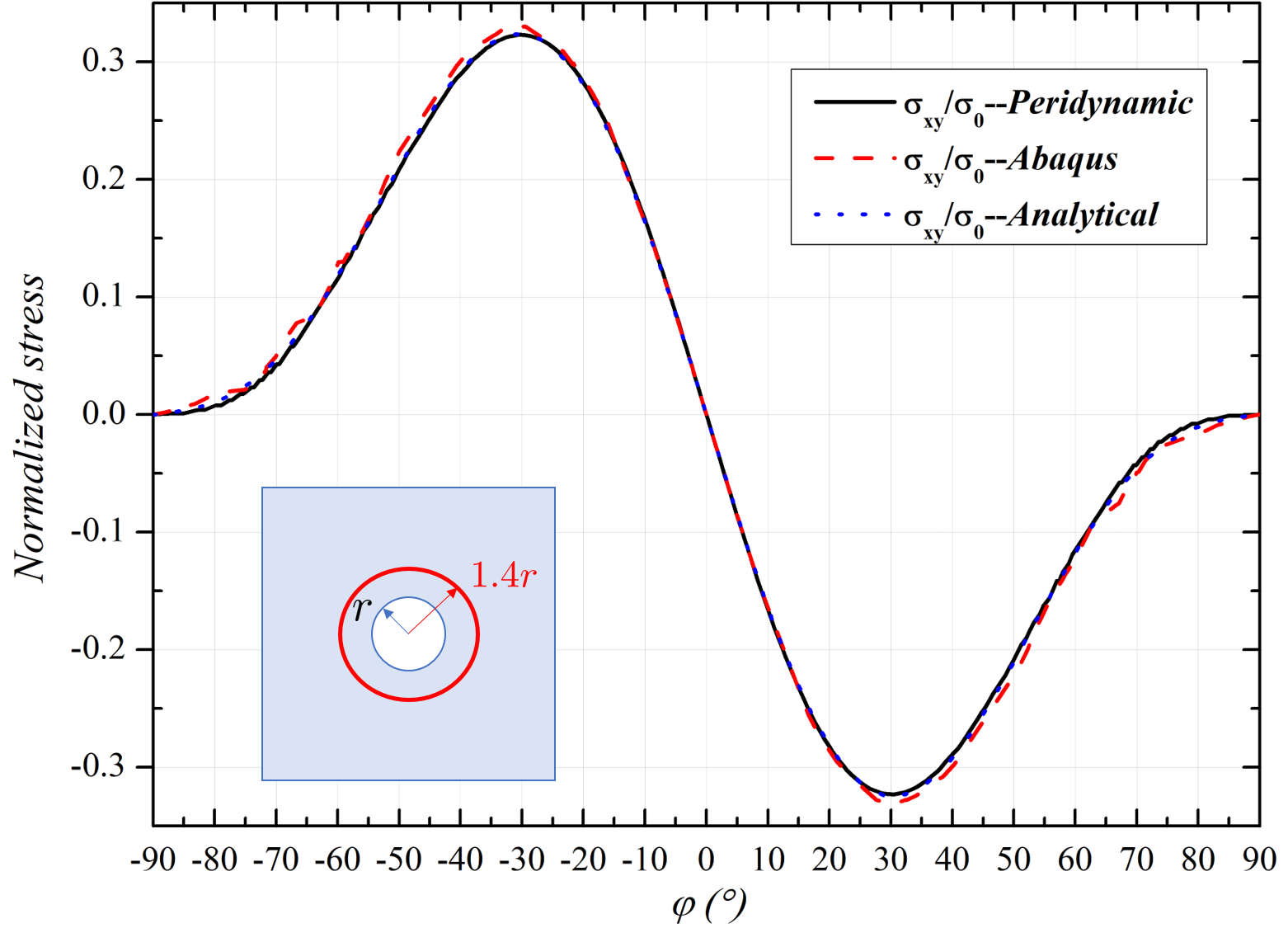}
\caption{
The normalized stresses $\sigma_{xy}/\sigma_0$ along the circular path with radius $R = 1.4r$ shown with a red circle.}
\label{fig:example1-7}
\end{figure}

In order to better illustrate the accuracy of the derived peridynamic stress tensor,
the normal stress $\sigma_{xx}$, $\sigma_{yy}$ are plotted along the horizontal ($y=0$mm)
and vertical ($x=0$mm) directions of the plate, while the shear stress $\sigma_{xy}$
is plotted along the circular path with radius $R=1.4r$,
as displayed in Figs. \ref{fig:example1-5} (a) and (b),
and Fig. \ref{fig:example1-7}, respectively.
As it can be seen from the figures,
the calculated peridynamic stresses show good agreements with finite element analysis results and analytical solutions.
While since the analytical solution is under the assumption of a circular hole in an elastic infinite medium,
there is a relatively small deviation in peridynamic and Abaqus results relative to that of analytical solutions,
which arises from
the finite dimensions of the problem considered in numerical simulations.
Meanwhile, as displayed in Fig. \ref{fig:example1-5},
a stress concentration factor $\omega$ predicted by the bond-based peridynamic theory
and finite element analysis is 2.98 and 2.96, respectively, which is consistent
with the analytical result of 3.0.

In order to clearly illustrate the stress concentration captured by the bond-based peridynamic stress formulation,
we zoomed in the region of vicinity of the hole in the plate,
as shown in Figs. \ref{fig:example1-8}-{\ref{fig:example1-10}.
It clearly shows the stress concentration
near the hole captured by the bond-based peridynamic model.

\begin{figure}[H]
\centering
\includegraphics[width=4.0in]{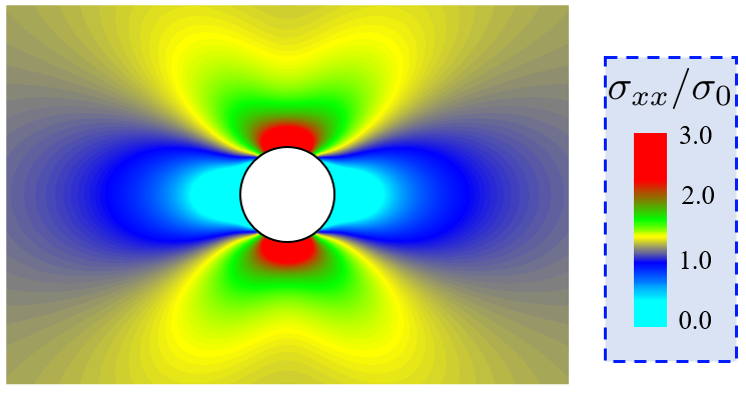}
\caption{
The normalized stress field $\sigma_{xx}/\sigma_0$ near the hole of the plate.}
\label{fig:example1-8}
\end{figure}
\begin{figure}[H]
\centering
\includegraphics[width=3.0in]{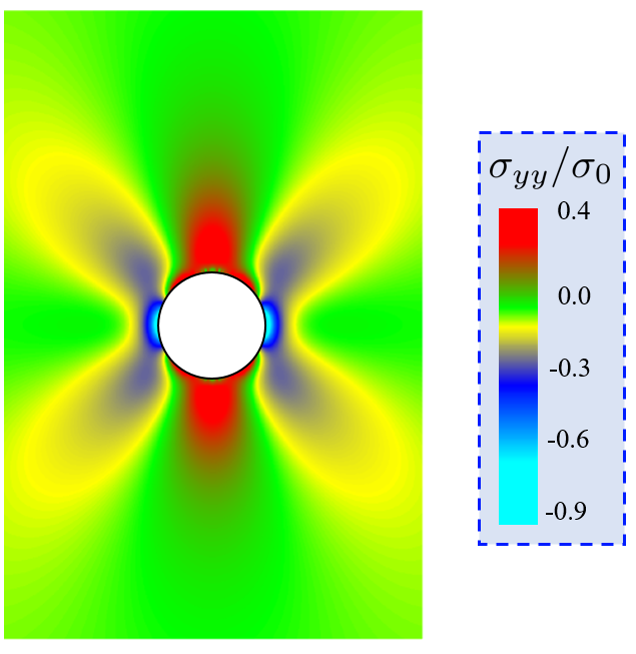}
\caption{
The normalized stress field $\sigma_{yy}/\sigma_0$ near the hole of the plate.}
\label{fig:example1-9}
\end{figure}
\begin{figure}[H]
\centering
\includegraphics[width=3.0in]{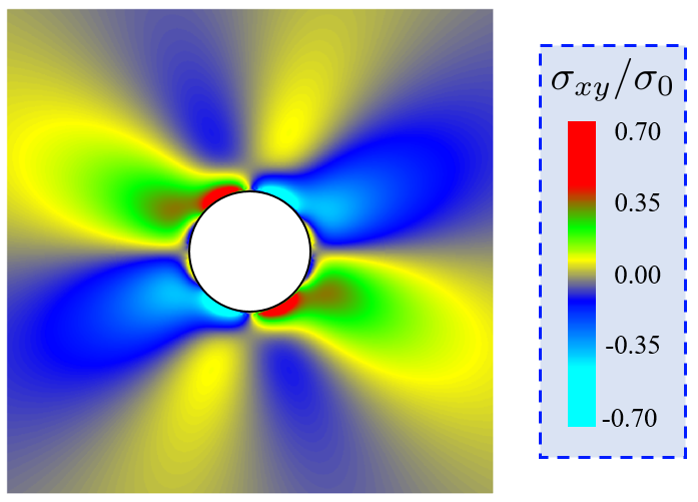}
\caption{
The normalized stress field $\sigma_{xy}/\sigma_0$ near the hole of the plate.}
\label{fig:example1-10}
\end{figure}

\subsection{A plate with a central crack under tension}

Next, we will consider an example of a square plate with a pre-existing crack at the center.
As shown in Fig. \ref{fig:example2-1},
the geometric parameters and material properties
are the same as in the first example of the plate with a hole.
The length of the crack is $a=10$mm.
While a tension load of $\sigma_0=10$MPa
is applied on both the top and the bottom edges of the plate.
In the peridynamic model,
the plate is discretized into about 0.72 million particles in a square arrangement with
the grid spacing of $\Delta = 0.059$mm and the horizon radius of $\delta = 3.015\Delta$.
The pre-existing crack is introduced by breaking
the bonds that cross the crack prior to the simulations.
A finite element analysis of the plate with a central crack was also performed to establish a baseline for the numerical results.
In the finite element analysis model,
the plate is meshed using 23686 bilinear plane stress quadrilateral elements leading to a total 24024 nodes.

\begin{figure}[H]
\centering
\includegraphics[width=3.0in]{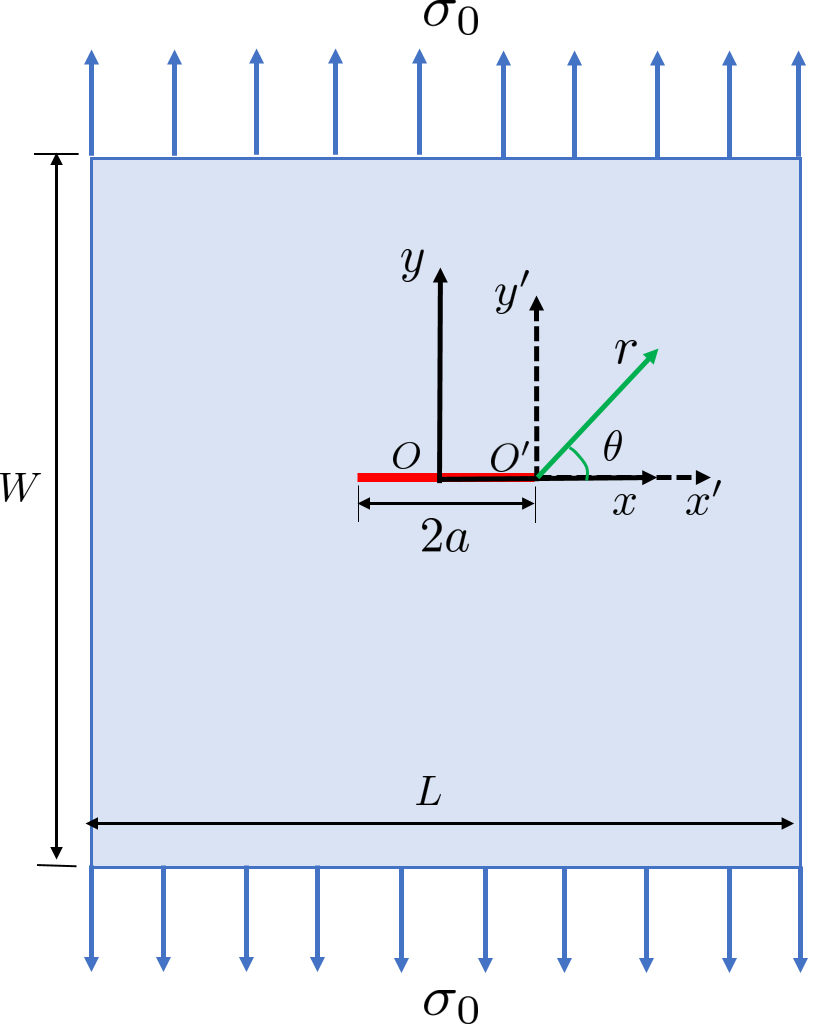}
\caption{
Graphic illustration of
the plate with a central crack loaded under uniaxial tension.}
\label{fig:example2-1}
\end{figure}

Figures \ref{fig:example2-2}-\ref{fig:example2-4} show the calculated
stress fields in the region near the central crack.
The stress distributions calculated from the proposed peridynamic formulation
agree well with those in finite element analysis.
In order to have a better comparison,
we also plotted the peridynamic stresses
along the horizontal ($y=0$mm) and vertical ($x=0$mm) directions,
and compared with that from finite element analysis,
as shown in Figs. \ref{fig:example2-5} (a) and (b), respectively.
Generally speaking,
the agreement between peridynamic and finite element analysis results
is good
despite the small deviation near the crack tip arising from the boundary effect.

\begin{figure}[H]
\centering
\includegraphics[width=5.0in]{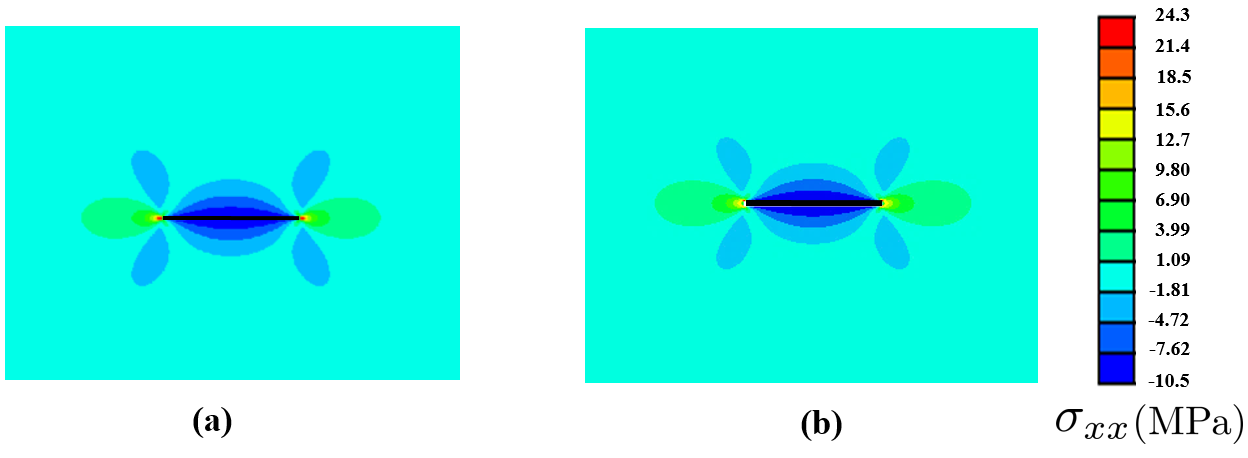}
\caption{
The normal stress $\sigma_{xx}$ for a plate with a central crack under tension
(a) bond-based peridynamic, (b) Abaqus results.}
\label{fig:example2-2}
\end{figure}
\begin{figure}[H]
\centering
\includegraphics[width=5.0in]{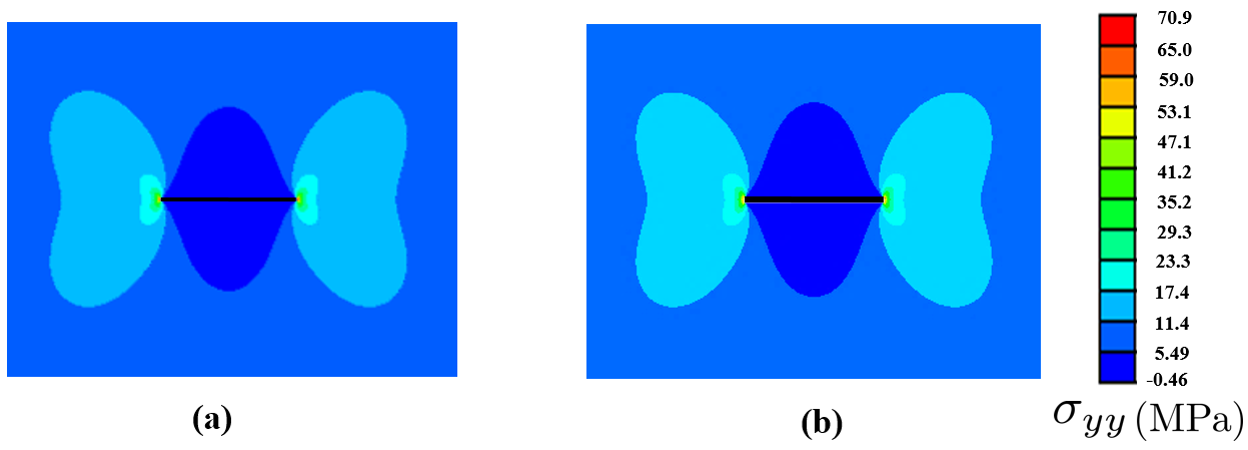}
\caption{
The normal stress $\sigma_{yy}$ for a plate with a central under tension (a) bond-based peridynamic, (b) Abaqus results.}
\label{fig:example2-3}
\end{figure}
\begin{figure}[H]
\centering
\includegraphics[width=5.0in]{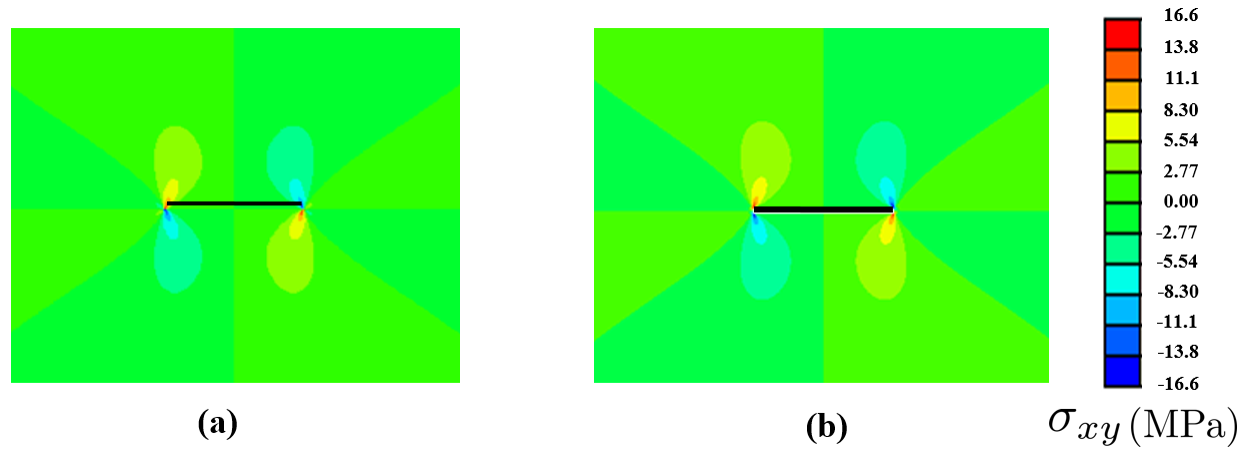}
\caption{
The shear stress $\sigma_{xy}$ for a plate with a central
under tension (a) bond-based peridynamic, (b) Abaqus results.}
\label{fig:example2-4}
\end{figure}
\begin{figure}[H]
\begin{minipage}{0.45\linewidth}
\begin{center}
\includegraphics[width=2.7in]{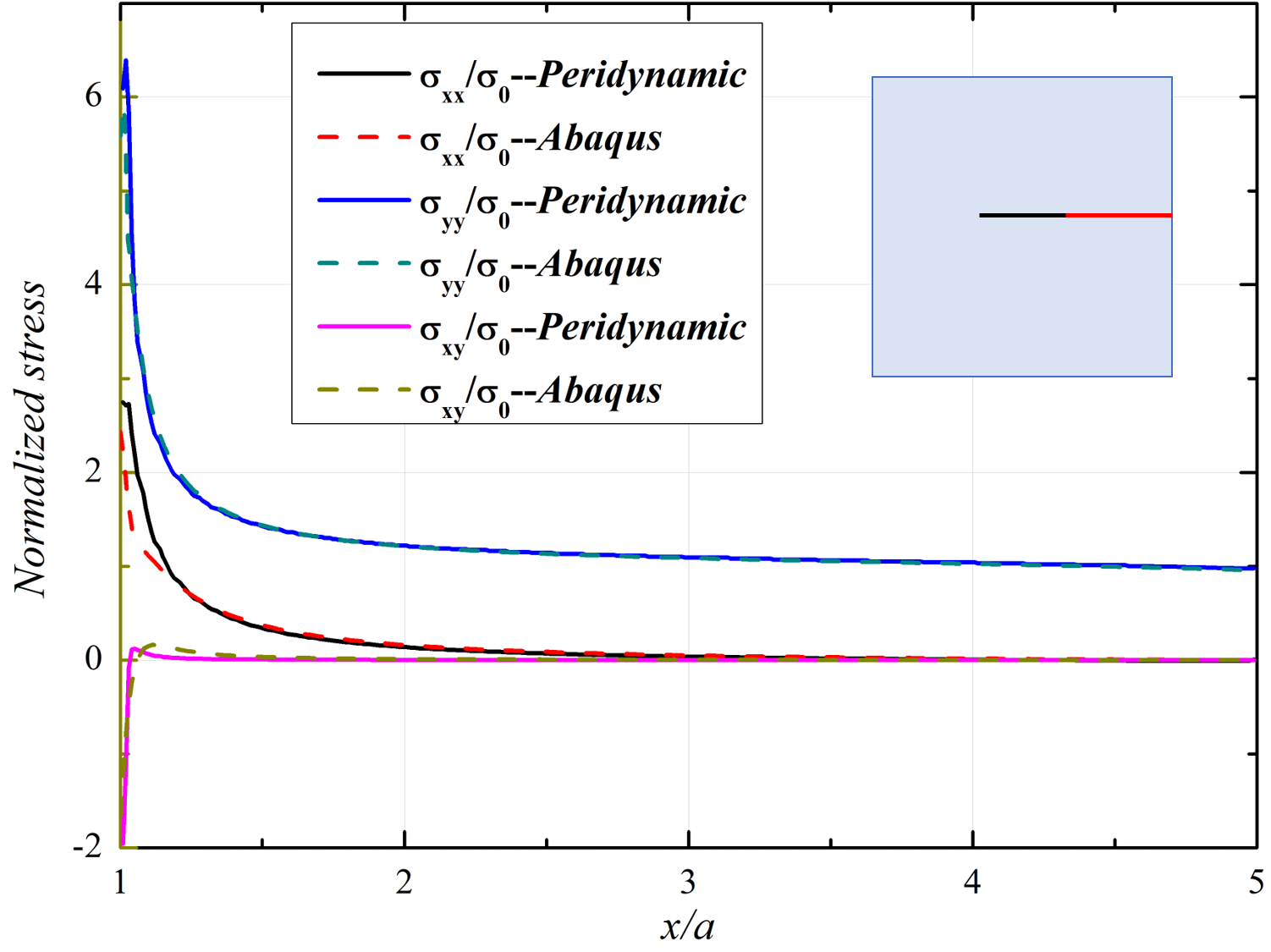}
\end{center}
\begin{center}
{(a)}
\end{center}
\end{minipage}
\begin{minipage}{0.45\linewidth}
\begin{center}
\includegraphics[width=2.7in]{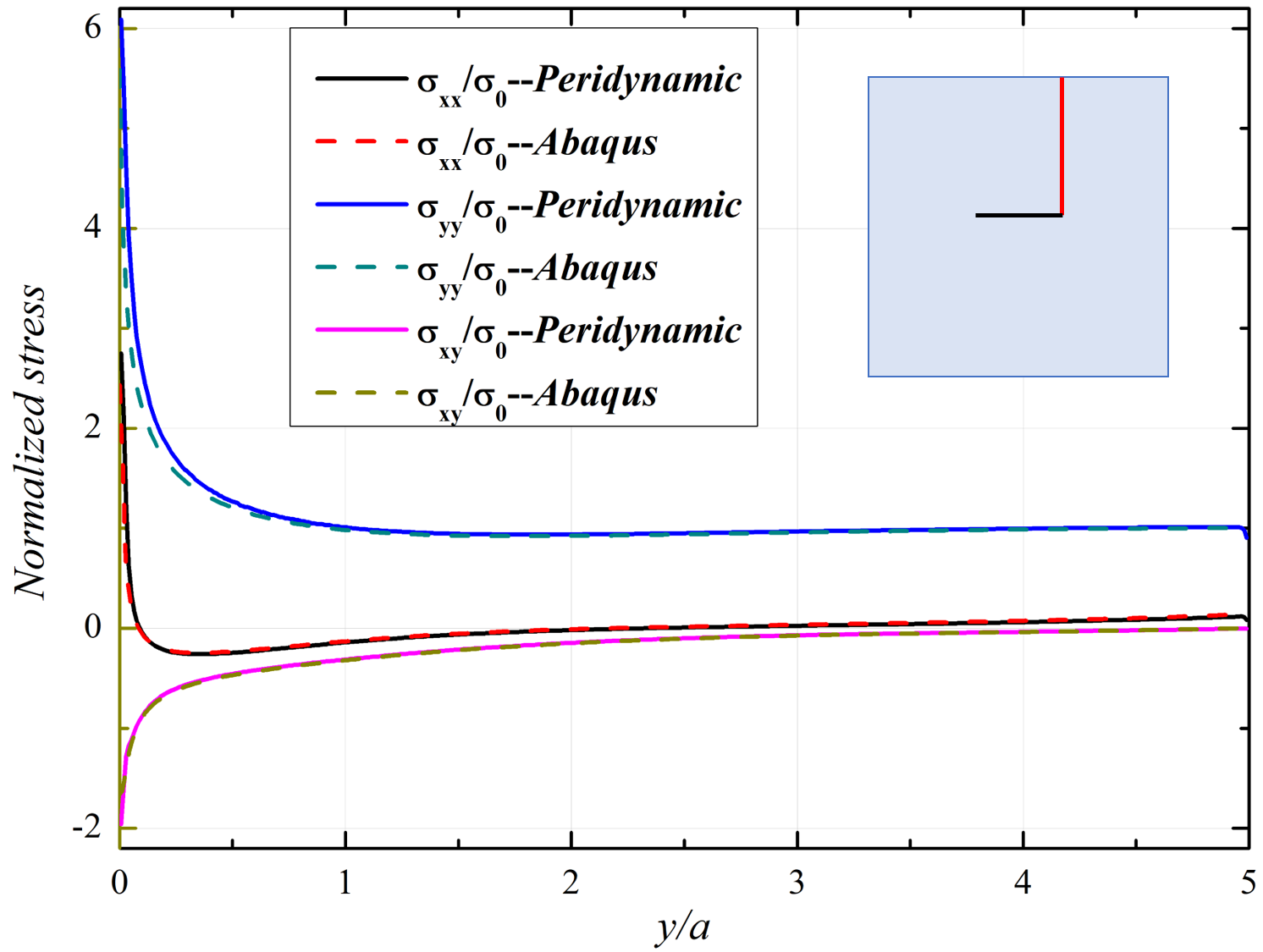}
\end{center}
\begin{center}
{(b)}
\end{center}
\end{minipage}
\caption{
(a)
The normalized stresses $\sigma_{xx}/\sigma_0$, $\sigma_{yy}/\sigma_0$ and $\sigma_{xy}/\sigma_0$ along the horizontal direction shown with a red line,
and (b) the normalized stresses $\sigma_{xx}/\sigma_0$, $\sigma_{yy}/\sigma_0$ and $\sigma_{xy}/\sigma_0$
along the vertical direction.}
\label{fig:example2-5}
\end{figure}

In order to describe the stress field around the crack tip region,
we then calculated the stress intensity factor (SIF).
The SIF was first introduced by Irwin \cite{Irwin1957}
to predict the stress state near the tip of
a crack caused by a remote load.
Since the fracture mode
of a plate with a central crack is crack opening, .ie. Mode I,
the stress field around crack tip under Mode I loading condition
for linear elastic materials can be written as \cite{Irwin1957},
\begin{eqnarray}
\sigma_{xx} &=& {K_I \over \sqrt{2 \pi r} } cos{\theta \over 2}
(1 - sin{\theta \over 2} sin{{ 3 \theta} \over 2})
\nonumber
\\
\sigma_{yy} &=& {K_I \over \sqrt{2 \pi r} } cos{\theta \over 2}
(1 + sin{\theta \over 2} sin{{ 3 \theta} \over 2})
\nonumber
\\
\sigma_{xy} &=& {K_I \over \sqrt{2 \pi r} } sin{\theta \over 2}
cos{\theta \over 2} cos{{ 3 \theta} \over 2}
\nonumber
\\
\sigma_{yz} &=& \sigma_{xz} = 0
\nonumber
\\
\sigma_{zz}
&=& \left \{
\begin{array}{lcl}
0, && {\rm plane~stress~condition}
\\
\\
{\nu}(\sigma_{xx} + \sigma_{yy}) ,  && {\rm plane~strain~condition}
\end{array}
\right .
\label{eq:crack-stress}
\end{eqnarray}
where $r$ an $\theta$ are the coordinates in the local cylindrical coordinate system at the crack tip (see Fig. \ref{fig:example2-1}).
$K_I$ is the stress intensity factor under Mode I loading condition,
which can be computed by the following theoretical equation \cite{Rooke1976},
\begin{eqnarray}
K_I = p\sqrt{\pi a}\Big[
{{1 - {a \over L} + 0.326({2a \over L})^2}  \over \sqrt{1-{2a\over L}} }
\Big].
\label{eq:SIF-Ana}
\end{eqnarray}

In the peridynamic model,
we can obtain
the $K_I$ by using displacement extrapolation method
proposed by Zhu and Oterkus \cite{Zhu2020},
\begin{eqnarray}
K_I = \sqrt{2\pi }{G \over {1+k}}{|\Delta v| \over \sqrt{r}},
\end{eqnarray}
where
$G$ is the shear modulus,
$k={{3-\nu}\over{1+\nu}}$ for plane stress condition and $k={3 - 4 \nu}$ for plane strain condition, $\Delta v$ is the relative displacement of one crack face with respect to the other,
$r$ the is coordinate in the local cylindrical coordinate system.
In the displacement extrapolation method,
the ${|\Delta v| \over \sqrt{r}}$ is assumed to be a linear function for the material point at the crack surface, .ie.
\begin{equation}
{|\Delta v| \over \sqrt{r}}
=a_1 + a_2 \cdot r.
\end{equation}
The unknown constants $a_1$ and $a_2$ can be determined by the
displacements of the selected material particles at the crack surface.
Since $$\lim_{r \to 0}{|\Delta v| \over \sqrt{r}} = a_1$$ at the crack tip,
the stress intensity factor $K_I$ within peridynamic framework can be computed as,
\begin{equation}
K_I = \sqrt{2\pi}{{Ga_1}\over{1+k}}.
\label{eq:SIF-PD}
\end{equation}
According to the Eq. (\ref{eq:SIF-PD}),
the calculated normalized SIF $K_I \over {p\sqrt{\pi a }}$ from bond-based peridynamics is 1.052, which agrees well with the analytical value of 1.021 (see Eq. \ref{eq:SIF-Ana}).

To enhance clarity,
the normalized stress field ${\sigma_{xx}\sqrt{2 \pi a}} / K_I$,
${\sigma_{yy}\sqrt{2 \pi a}} / K_I$ ,
and ${\sigma_{xy}\sqrt{2 \pi a}} / K_I$
in the vicinity of the crack-tip is presented in Fig. \ref{fig:example2-7}(a),
Fig. \ref{fig:example2-8}(a), and Fig. \ref{fig:example2-9}(a), respectively.
As it can be seen from the figures,
the peridynamic stress formulation
is accurate that can be employed to describe
the regions of stress concentration
where crack initiation is likely to occur.
To make a better comparison with
linear elastic analysis,
we also plotted the angular variations of the normalized stress ${\sigma_{xx}\sqrt{2 \pi r}} / K_I$, ${\sigma_{yy}\sqrt{2 \pi r}} / K_I$,
and ${\sigma_{xy}\sqrt{2 \pi r}} / K_I$, around the crack tip for rings of different radii and compared with
the solution based on linear elastic fracture mechanics (see \ref{eq:crack-stress}),
as shown in Fig. \ref{fig:example2-7}(b),
Fig. \ref{fig:example2-8}(b), and Fig. \ref{fig:example2-9}(b) respectively.
The figures indicate that
the calculated peridynamic stresses show good agreements with
linear elastic results when the value of $r$ is greater than the size of horizon.
In contract,
if the value of $r$ is smaller than that of horizon,
the estimated peridynamic stresses
differ greatly from the linear elastic solutions.
While as shown in Fig. \ref{fig:example2-7}(b) and Fig. \ref{fig:example2-8}(b),
although the calculated peridynamic
stresses ${\sigma_{xx}\sqrt{2 \pi r}} / K_I$ and ${\sigma_{yy}\sqrt{2 \pi r}} / K_I$
exhibit the similar variation trend with those
of the linear elastic analysis, there are some deviations
especially
near the crack tip and behind the crack front, which mainly arises from the boundary effect.

\begin{figure}[H]
\begin{minipage}{0.45\linewidth}
\begin{center}
\includegraphics[width=2.5in]{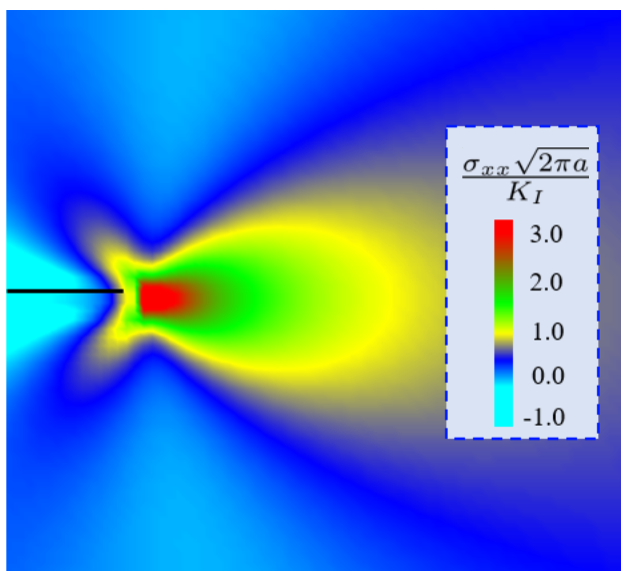}
\end{center}
\begin{center}
{(a)}
\end{center}
\end{minipage}
\begin{minipage}{0.45\linewidth}
\begin{center}
\includegraphics[width=2.8in]{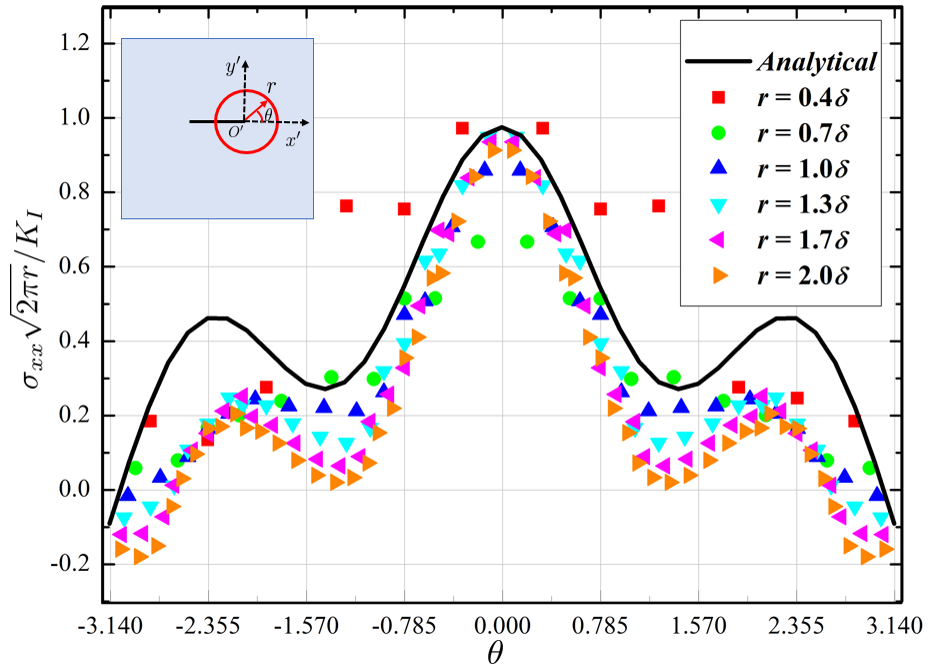}
\end{center}
\begin{center}
{(b)}
\end{center}
\end{minipage}
\caption{
(a)
The normalized stress field ${\sigma_{xx}\sqrt{2 \pi a}} / K_I$ in the vicinity of the crack tip,
and (b) the angular variations of the normalized stress ${\sigma_{xx}\sqrt{2 \pi r}} / K_I$  around the crack tip for rings of different radii.}
\label{fig:example2-7}
\end{figure}
\begin{figure}[H]
\begin{minipage}{0.45\linewidth}
\begin{center}
\includegraphics[width=2.5in]{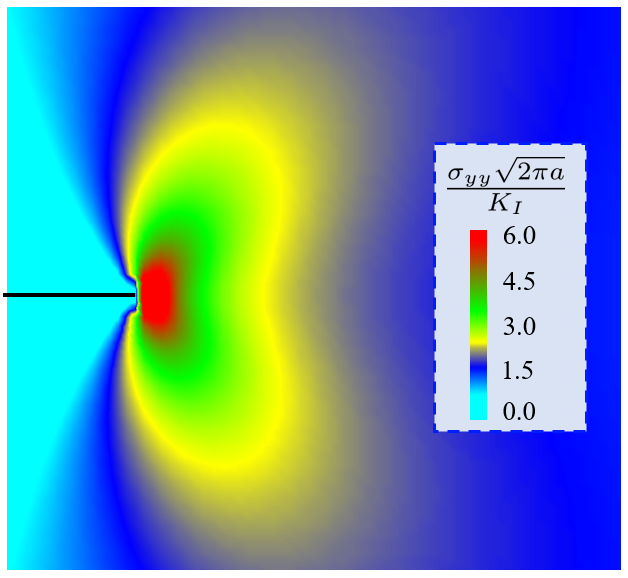}
\end{center}
\begin{center}
{(a)}
\end{center}
\end{minipage}
\begin{minipage}{0.45\linewidth}
\begin{center}
\includegraphics[width=2.8in]{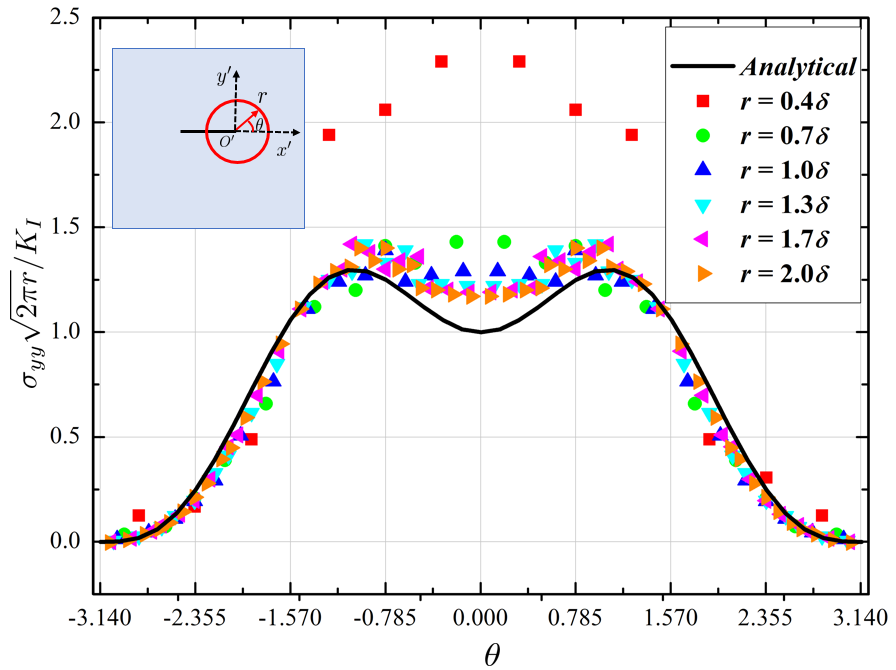}
\end{center}
\begin{center}
{(b)}
\end{center}
\end{minipage}
\caption{
(a)
The normalized stress field ${\sigma_{yy}\sqrt{2 \pi a}} / K_I$ in the vicinity of the crack tip,
and (b) the angular variations of the normalized stress ${\sigma_{yy}\sqrt{2 \pi r}} / K_I$  around the crack tip for rings of different radii.}
\label{fig:example2-8}
\end{figure}
\begin{figure}[H]
\begin{minipage}{0.45\linewidth}
\begin{center}
\includegraphics[width=2.5in]{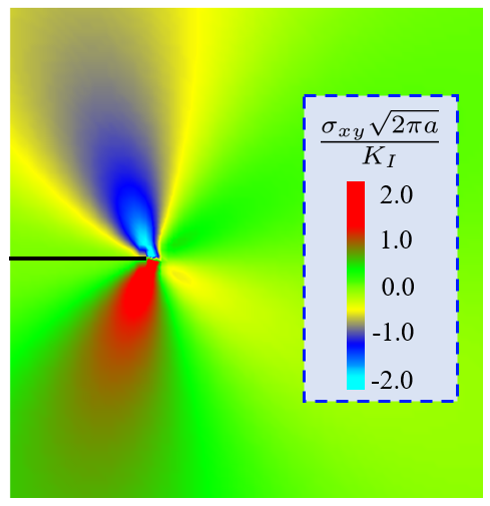}
\end{center}
\begin{center}
{(a)}
\end{center}
\end{minipage}
\begin{minipage}{0.45\linewidth}
\begin{center}
\includegraphics[width=2.8in]{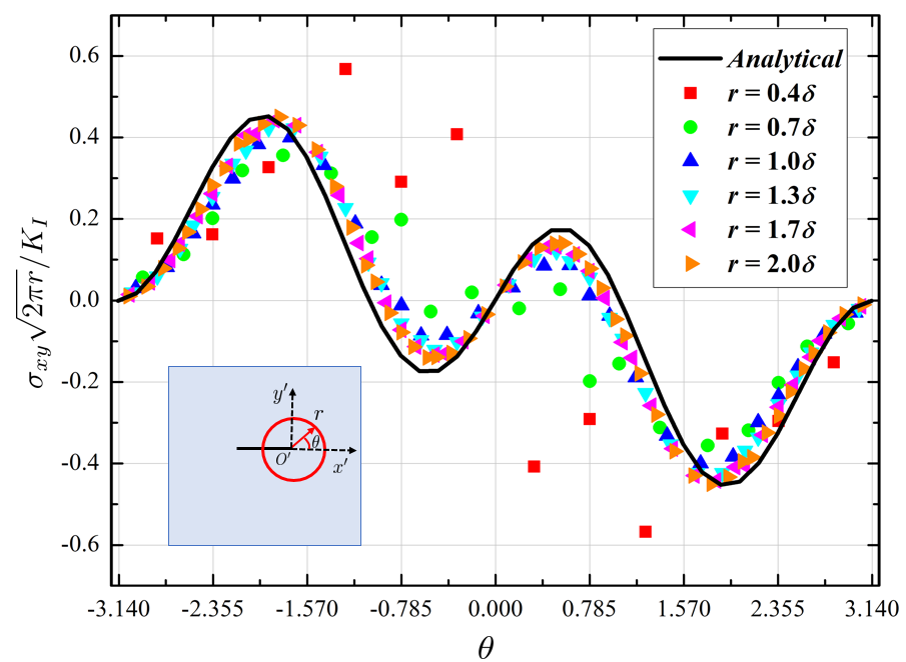}
\end{center}
\begin{center}
{(b)}
\end{center}
\end{minipage}
\caption{
(a)
The normalized stress field ${\sigma_{xy}\sqrt{2 \pi a}} / K_I$ in the vicinity of the crack tip,
and (b) the angular variations of the normalized stress ${\sigma_{xy}\sqrt{2 \pi r}} / K_I$  around the crack tip for rings of different radii.}
\label{fig:example2-9}
\end{figure}

To compare with the peridynamic crack solution with
the analytical asymptotic solution at crack tip,
we first calculate
the Von Mises effective stress $\sigma_e$ at the crack tip, which
is expressed as,
\begin{equation}
\sigma_e^2 = { 1 \over 2}
[(\sigma_{xx} - \sigma_{yy})^2 + (\sigma_{yy} - \sigma_{zz})^2
+ (\sigma_{zz} - \sigma_{xx})^2]
+ 3(\sigma_{xy}^2 + \sigma_{xz}^2 + \sigma_{yz}^2),
\label{eq:von-Mises}
\end{equation}
where $\sigma_{ij}$ is the stress components in the Cartesian coordinate system.
For a fixed $\sigma_e$, we can find the constant $\sigma_e$ contour $r(\theta)$
by substituting Eq. (\ref{eq:crack-stress}) into Eq. (\ref{eq:von-Mises}),
\begin{eqnarray}
r(\theta) =
\left \{
\begin{array}{lcl}
\displaystyle
{ 1 \over {4 \pi}} ({K_I \over \sigma_e})^2
[( 1 + cos \theta) + {3 \over 2} sin^2 \theta], && {\rm plane~stress~condition}
\\
\\
\displaystyle
{ 1 \over {4 \pi}} ({K_I \over \sigma_e})^2
[(1 - 2\nu)^2 ( 1 + cos \theta) + {3 \over 2} sin^2 \theta] ,  && {\rm plane~strain~condition}
\end{array}
\right .
\end{eqnarray}
which is usually referred as the plastic zone contour.
This is because for small scale yield the interior area of $r(\theta)$ contour
may be viewed as the plastic zone size or shape.

Figure \ref{fig:example2-10} displays the near-tip plastic zone shape and size estimated
by using bond-based peridynamics along with the solution of
linear elastic fracture mechanics, when the applied stress level
\[
K_I / (\sigma_e \sqrt{2 \pi a}) = 0.2~~\to~~ \sigma_e = {K_I \over 0.2 \sqrt{2\pi a}}~.
\]
In Fig. \ref{fig:example2-10}, we plot the radial distance of any point in
the contour of a contact $\sigma_e$ to the crack tip, i.e.  $r(\theta)$, as the function of
angle variation $\theta$, and we compare it with the solution of
LEFM
under plane stress condition.
For ease comparison,
all the sizes are normalized.
As shown in Fig. \ref{fig:example2-10},
there are some discrepancies between
the peridynamic solution and the solution based on linear elastic fracture mechanics (LEFM),
especially behind the crack front.
The boundary effect in peridynamics may be the main reason for the discrepancy.
Another possible reason for the difference is that
the plastic zone estimated by LEFM is based on
the assumption that the material is linear elastic with infinitesimal deformation.
On the other hand,
for the peridynamic PMB model, as shown in previous section,
it corresponds to a nonlinear elastic constitutive behaviors at large deformations,
which may deviate from the results of LEFM, when the stress level becomes
sufficiently large.
It may be also possible that this discrepancy is due to
the particle density or resolution at the crack tip is not
enough, and it is believed that
the bond-based peridynamic stress solution may be
improved by employing adaptive refinement \cite{Warren2009} in the vicinity of the crack tip.

\begin{figure}[H]
\centering
\includegraphics[width=4.0in]{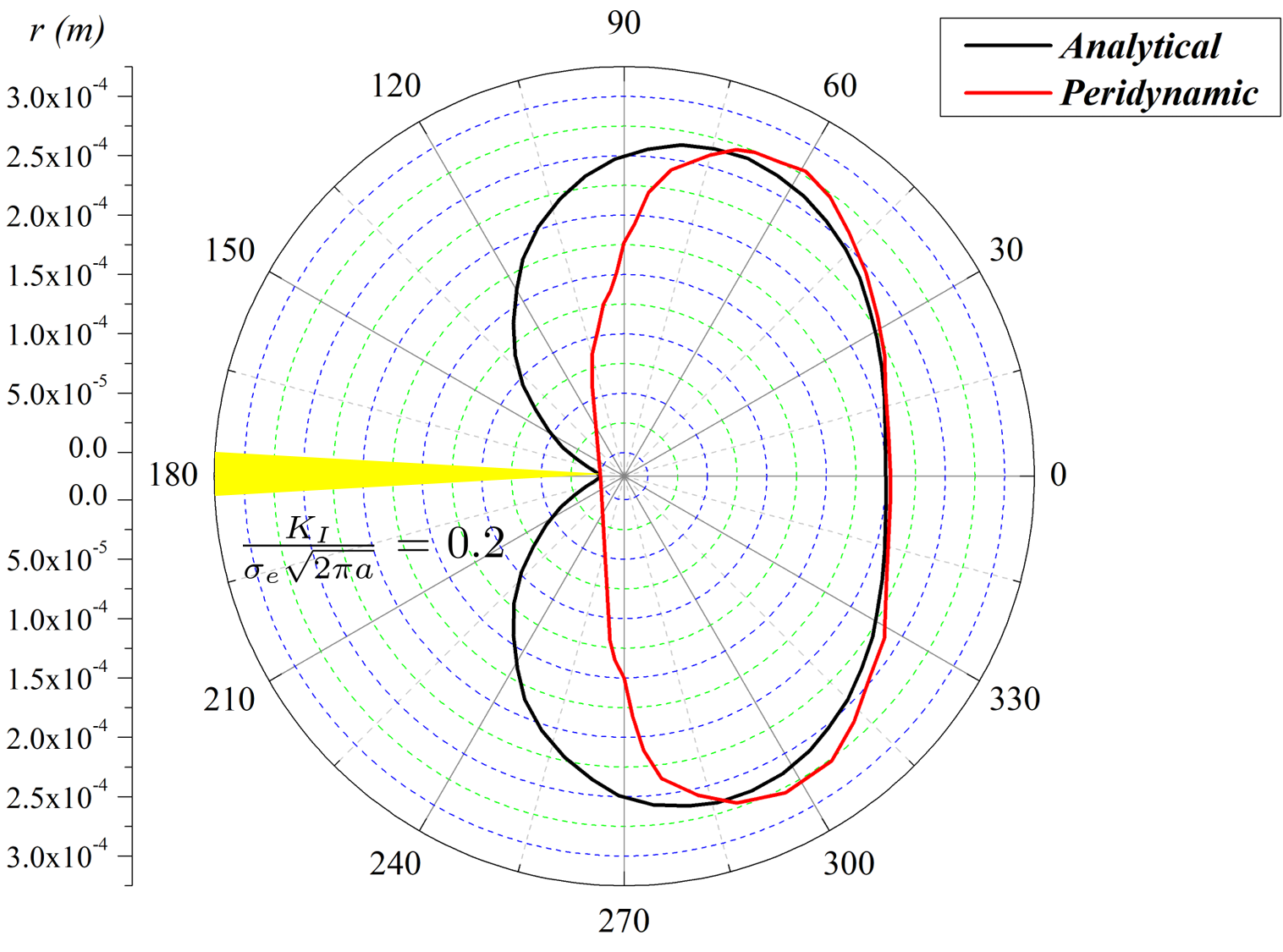}
\caption{
 the crack tip plastic zone obtained by linear elastic estimate and peridynamic analysis.}
\label{fig:example2-10}
\end{figure}

\subsection{Three-dimensional double edge notch specimen under tension}

The last example
is to evaluate 3-dimensional stress concentrations around notches
using derived bond-based peridynamic stress formulation.
Here, we considered an example of a homogeneous tensile specimen with double edge notches
in the middle section, as shown in Fig. \ref{fig:example3-1}.
In experimental studies \cite{Pindera1992},
strain gages were placed at key locations near the notches,
and information on the strain/stress stated was obtained using miniature strip electric resistance gages having a pitch between 2 and 3mm.
The specimen is made of aluminum alloy 2024-T4 with
$E=73$GPa and $\nu=0.32$.
 Due to the symmetry of the problem, only one-quarter of the specimen is modeled in the peridynamic model.
Fig. \ref{fig:example3-1} shows the geometric parameters of the specimen where a tension load $\sigma_0=14.5$MPa is applied on the top, and the bottom is placed on rollers.
In the peridynamic model.
the geometry of the specimen is discretized using about 1.5 million particles in a hexagonal arrangement with a horizon size of 3.015 times the grid spacing of 0.5 mm.
While in the finite element analysis model,
only an eighth of the specimen was modeled using 24588 10-node tetrahedral elements by considering symmetry.

\begin{figure}[H]
\centering
\includegraphics[width=2.5in]{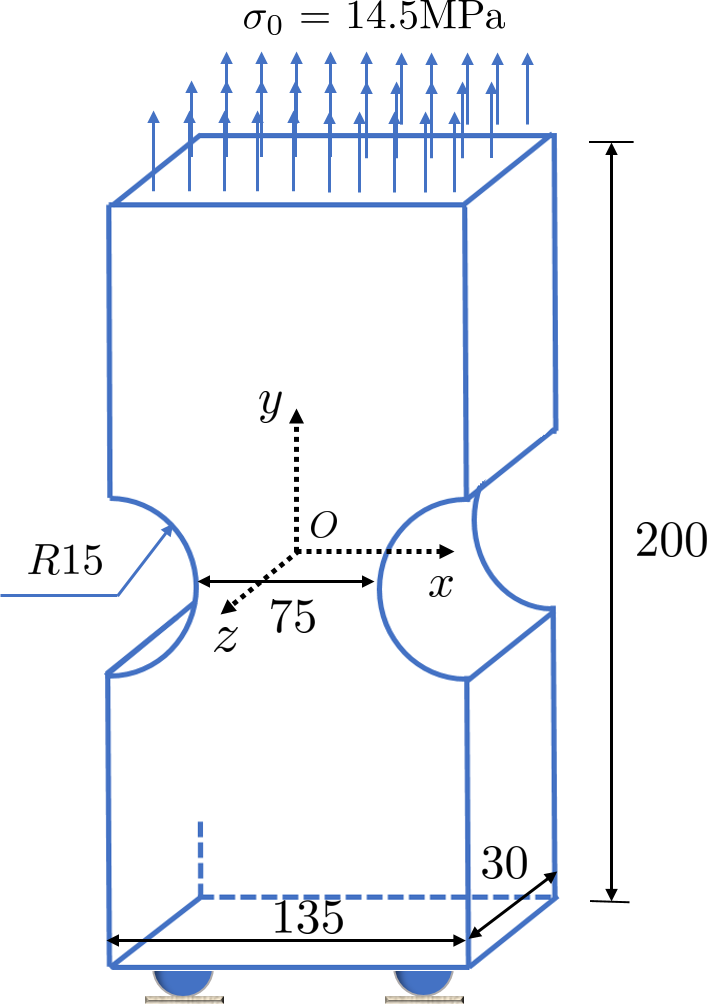}
\caption{
Graphic illustration of
2024-T4 aluminum alloy specimen with double edge notches loaded under uniaxial tension (unit=mm).}
\label{fig:example3-1}
\end{figure}

Figs. \ref{fig:example3-2} and \ref{fig:example3-3}
show the calculated bond-based peridynamic stress distributions along the face ahead of the notch ($y=0$mm, $z=15$mm) and along the base of the notch ($x=22.5$mm, $y=0$mm), respectively,
and compared with that of finite element analysis, experimental data, as well as non-ordinary state-based peridynamics \cite{Breitenfeld2014},
to illustrate the accuracy of proposed peridynamic stress formulation in dealing with 3-dimensional problems.
Along the face of the notch, as shown in Fig. \ref{fig:example3-2},
the numerical solutions and experimental data show excellent agreements,
which suggest that
the proposed peridynamic stress formula enables
to evaluate the stress concentration at the notch
within the framework of bond-based peridynamics.
While along the base of the notch,
although the calculated peridynamic stresses exhibit similar
increasing trend from the outer edge to the middle point with that of the experiment,
the stress values are lower than those measured in the experiment.
There are several reasons for the discrepancy.
One possible reason is that
the curvature is not entirely smooth in the experiment,
while the base of the notch is flat.
Another reason is that since the bond-based peridynamics is constrained
with a fixed Poisson's ratio of $1/4$ for 3-dimensional conditions,
it causes some deviations
in describing the material constitutive properties, thereby causing
the discrepancy in stresses.
One possible way to improve the
bond-based peridynamic stress solutions
is to capture the full curvature
employing adaptive refinement \cite{Warren2009} around the notch.

\begin{figure}[H]
\centering
\includegraphics[width=4.0in]{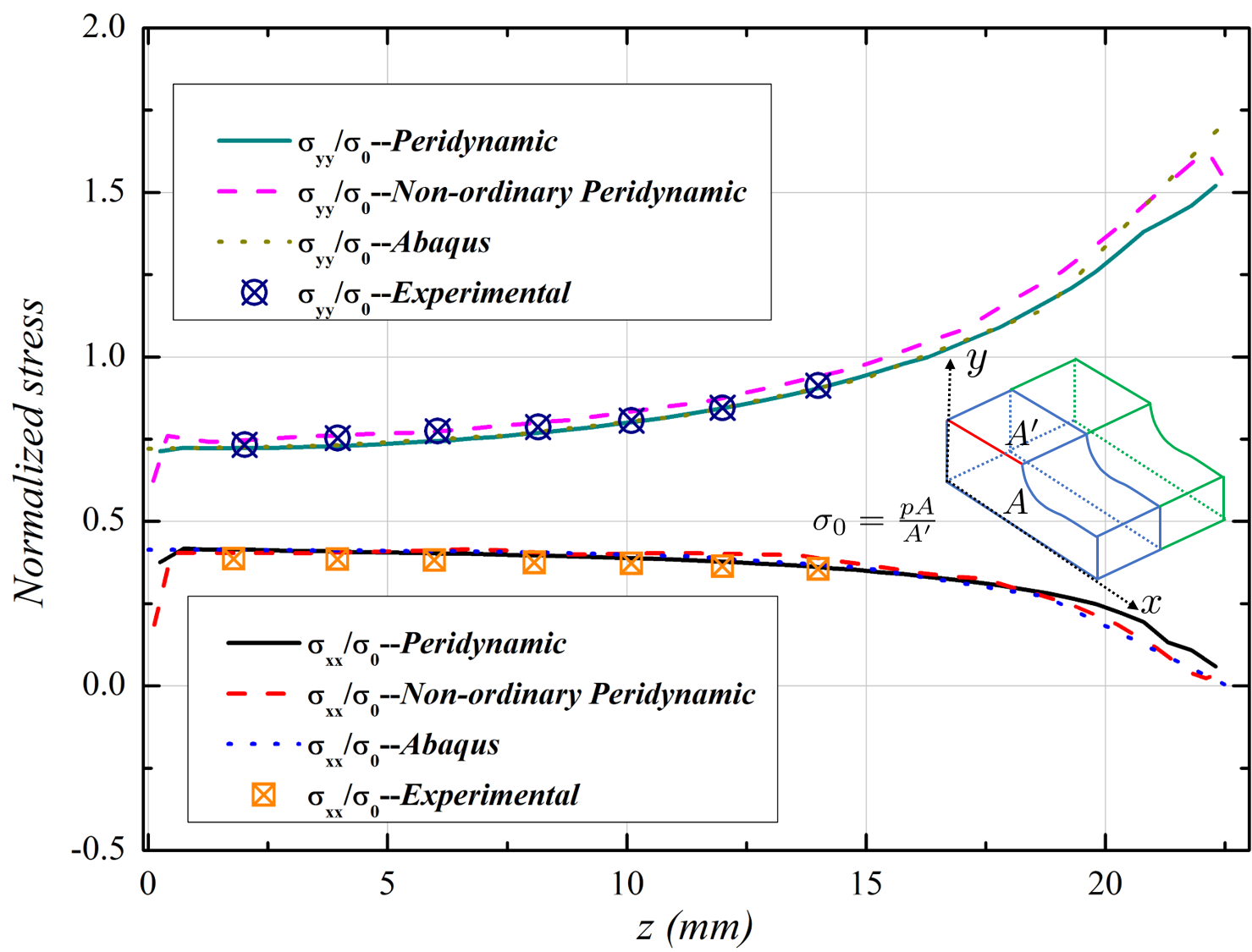}
\caption{
The normalized stress field $\sigma_{xx} / \sigma_{0}$ and $\sigma_{yy} / \sigma_{0}$ along the face ahead of the notch, where the non-ordinary state-based peridynamic data is from
Ref. \cite{Breitenfeld2014}.}
\label{fig:example3-2}
\end{figure}
\begin{figure}[H]
\centering
\includegraphics[width=4.0in]{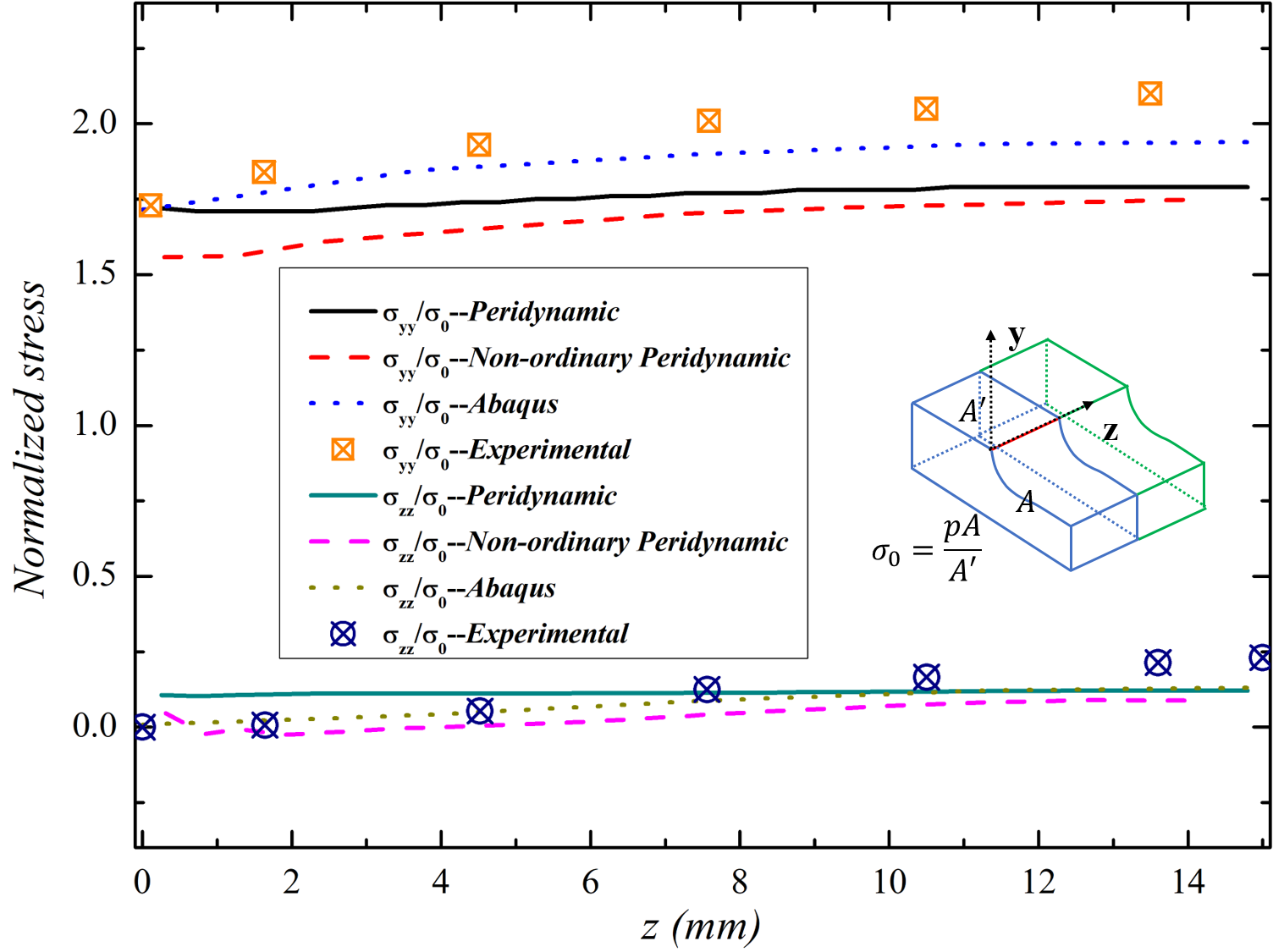}
\caption{
The normalized stress field $\sigma_{yy} /\sigma_{0}$ and $\sigma_{zz} / \sigma_{0}$ along the base of the notch, where the non-ordinary state-based peridynamic data is from Ref. \cite{Breitenfeld2014}.}
\label{fig:example3-3}
\end{figure}

In order to show the stress concentration clearly,
we zoomed in the region around the notch,
as shown in Figs. \ref{fig:example3-4}-{\ref{fig:example3-6}.
The calculated peridynamic stress distributions
clearly show that
the stress concentration occurs around region of the notch.
It means that
the bond-based peridynamics can be employed to
capture regions of stress concentration using
the derived peridynamic stress formulation.

\begin{figure}[H]
\centering
\includegraphics[width=3.0in]{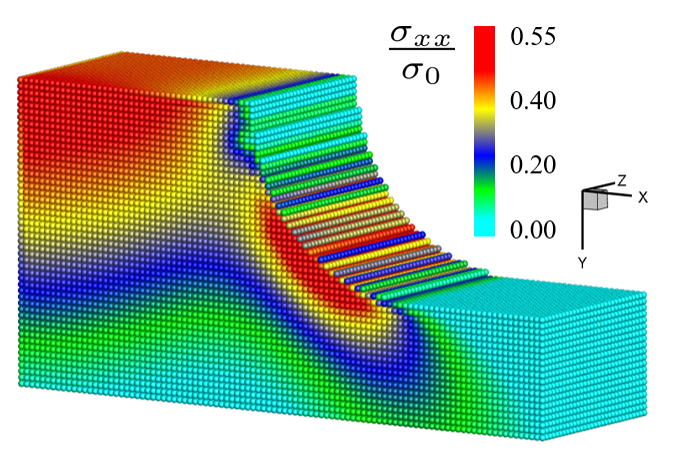}
\caption{
The normalized stress field $\sigma_{xx} / \sigma_0$  in the vicinity of the notch.}
\label{fig:example3-4}
\end{figure}
\begin{figure}[H]
\centering
\includegraphics[width=3.0in]{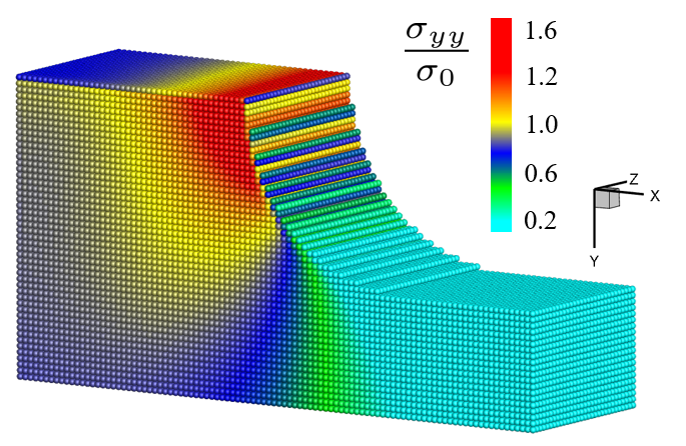}
\caption{
The normalized stress field $\sigma_{yy} / \sigma_0$  in the vicinity of the notch.}
\label{fig:example3-5}
\end{figure}
\begin{figure}[H]
\centering
\includegraphics[width=3.0in]{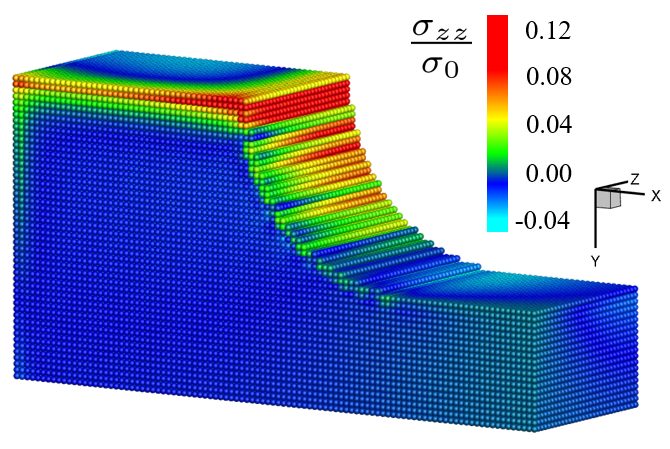}
\caption{
The normalized stress field $\sigma_{zz} / \sigma_0$  in the vicinity of the notch.}
\label{fig:example3-6}
\end{figure}

\section{Conclusions}
\label{sec:Conclusion}

In the present work,
we have shown that
the peridynamic stress
is exactly the same as the
first Piola-Kirchhoff Virial stress
which stemmed from Irving-Kirkwood-Noll formalism
and the Hardy-Murdoch procedure.
The expression of developed
peridynamic stress formula is
much simple that can be easily implemented in numerical simulations.
We then applied the
peridynamic stress formulation
to simulate both 2-dimensional and 3-dimensional problems
considering singularity and discontinuities.
The peridynamic stress is evaluated within the bond-based peridynamics using PMB material model.
It is found that
the PMB model may exhibit nonlinear constitutive
behaviors to demonstrate material geometric nonlinearity at large deformations.
Emphasis is placed on evaluating the
accuracy of the peridynamic stress in the region of stress concentrations, involving the hole, the crack tip,
and notches.
The accuracy of the peridynamic stress
is verified
by comparing with finite element analysis results, analytical solutions,
and experimental data with good agreements.

The developed peridynamic stress formulation provides the ability
to predict stress distributions within the framework of
bond-based peridynamics.
As a consequence,
the bond-based peridynamics
can describe the constitutive behaviors of a material in terms of a stress tensor rather than just the bond force.
Thus,
the bond-based peridynamics
is not only useful
for simulating crack propagation,
but also can be employed to capture regions of stress concentrations
during discontinuous deformations.

\section*{References}

\bibliography{mybibfile}
\end{document}